\documentclass[pdftex,twocolumn,epjc3]{svjour3}          

\RequirePackage[T1]{fontenc}

\smartqed  

\RequirePackage{graphicx}
\RequirePackage{newtxtext,newtxmath}
\RequirePackage{flushend}
\RequirePackage[numbers,sort&compress]{natbib}
\RequirePackage[colorlinks,citecolor=blue,urlcolor=blue,linkcolor=blue]{hyperref}
\RequirePackage{amsmath}
\RequirePackage{doi}
\RequirePackage{color}
\RequirePackage{placeins} 

\RequirePackage{xprintlen}
\RequirePackage[switch]{lineno}

\RequirePackage{url}

\journalname{Eur. Phys. J. C}

\definecolor{darkgreen}{rgb}{0.0,0.4,0.0}
\DeclareMathAlphabet\mathbfcal{OMS}{cmsy}{b}{n}

\DeclareMathOperator{\sign}{sign}

\begin{document}

\title{Bayesian inference of high-purity germanium detector impurities 
based on capacitance measurements and machine-learning accelerated capacitance calculations}


\author{I.~Abt\thanksref{addr1}\and
    C.~Gooch\thanksref{addr1}\and
    F.~Hagemann\thanksref{addr1}\and
    L.~Hauertmann\thanksref{e1,addr1}\and
    X.~Liu\thanksref{addr1}\and
    O.~Schulz\thanksref{addr1}\and
    M.~Schuster\thanksref{addr1}
}

\thankstext{e1}{e-mail: lhauert@mpp.mpg.de (corresponding author)}

\institute{Max Planck Institut für Physik, Föhringer Ring 6, 80805 Munich, Germany\label{addr1}
}

\date{Received: date / Accepted: date}

\maketitle



\begin{abstract}
The impurity density in high-purity germanium detectors
is crucial to understand and simulate such detectors.
However, the information about the impurities provided by the manufacturer,
based on Hall effect measurements, is typically limited
to a few locations and comes with a large uncertainty.
As the voltage dependence of the capacitance matrix of a detector
strongly depends on the impurity density distribution,
capacitance measurements can provide a path to improve the knowledge on the impurities. 
The novel method presented here uses a machine-learned surrogate model, 
trained on precise GPU-accelerated capacitance calculations, to perform 
full Bayesian inference of impurity distribution parameters from capacitance measurements.
All steps use open-source Julia software packages. 
Capacitances are calculated with \emph{SolidStateDetectors.jl}, machine learning 
is done with \emph{Flux.jl} and Bayesian inference performed using \emph{BAT.jl}.
The capacitance matrix of a detector and its dependence on the impurity density is explained
and a capacitance bias-voltage scan of an $n$-type true-coaxial test detector is presented.
The study indicates that the impurity density of the test detector also has a radial dependence.

\end{abstract}

\section{Introduction}
\label{sec:introduction}

Advanced scientific applications of high-purity germanium (HPGe) detectors often require a quantitative 
understanding on how the detector signal depends on the event topology. 
This requires a realistic simulation of the detector. Example applications are searches for 
physics beyond the Standard Model such as neutrinoless double-beta 
decay~\cite{LEGEND:2017cdu,0bnn:Review2019,Majorana:2019nbd,GERDA:2020xhi,LEGEND:2021bnm}
and dark matter~\cite{CoGeNT:2013,SuperCDMS:2014,PhysRev:LimitWIMPS2018}.
In such rare-event searches, it is essential to distinguish between signal 
and background events based on the shape of the detector signal.

A very important input to the simulation of 
HPGe detectors is the impurity density distribution, $\zeta$, 
of the electrically active impurities in the germanium crystal as it 
strongly influences the electric potential, $\Phi$. 
The calculation of $\Phi$ is the first step in the simulation of an HPGe detector.
Software packages like \emph{SolidStateDetectors.jl}~\cite{Abt:2021mzq} (SSD), 
MJDSigGen~\cite{Radford:siggen}
or the AGATA Detector Library~\cite{Bruyneel:2016zih}
are able to perform a full simulation of HPGe detectors from field calculation to signal formation. 
It is crucial, though, to use the correct $\zeta$ to obtain the correct electric field,
which influences the drift of the charge carriers and, thus, the formed pulses. 
Often, the impurity density is measured via the Hall effect at different heights of a drawn crystal 
ingot by cutting thin slices out of the ingot. However, there is a rather large uncertainty on 
these impurity measurements and, in addition, 
assumptions have to be made how the impurity density changes in between.
Usually, only a linear or quadratic change of the impurity density 
between the bottom and the top of a cylindrically shaped crystal is assumed. 
If an incorrect model for $\zeta$ is assumed, wrong conclusions can be drawn in 
studies involving the subsequent parts of the simulation, e.g. in studies involving the mobility tensor.

An HPGe detector is, in principle, a $p-n$ diode operated in reversed bias mode. 
The extent of the depleted volume for different bias voltages, $U_B$, depends on $\zeta$. 
The undepleted volumes are extensions of the detector contacts. 
Thus, the capacitance between those contacts depends on $U_B$ and $\zeta$. 
Therefore, $\zeta$ can be studied by measuring the capacitance for different $U_B$, a C-V curve, and comparing it to simulated C-V curves for different $\zeta$.

Past work by Bruyneel et al.~\cite{Bruyneel2011SRR, Birkenbach2011HPGEImp} has shown that it is indeed possible to determine impurity density parameters based on C-V measurements.
It was, however, limited to a best-fit approach without uncertainty estimates and with very few free parameters. 
Moving beyond this to impurity density models that have many free parameters and fully exploring such parameter spaces
is very challenging due to the high numerical cost: Simulating a C-V curve requires repeated and numerically expensive field calculations.
Even with the GPU-accelerated implementation in SSD and using multiple GPUs in parallel,
it takes a few minutes to calculate one C-V curve. It is, therefore, prohibitive to perform these calculations directly during parameter inference as a proper exploration of the parameter space would take a very long time. The novel method presented here circumvents this problem by replacing the exact capacitance calculations with a machine-learned approximation function. It comprises the following steps:
\begin{enumerate}
    \item Definition of a model for $\zeta$ for a detector and including the allowed parameter space for its parameters, $p$.
    \item Quasi-random generation of $N$ parameters sets: $X = \{p_i\}$ for $i \in [1,N]$.
    \item Calculation of the capacitance, $c_i$, via SSD for each element of $X$: $Y= \{c_i\}$ for $i \in [1,N]$.
    \item Training of a deep neural network, $\mathcal{DNN}$, 
        on the generated data set: ($X$|$Y$).
    \item Bayesian inference of the model on a measured C-V curve 
        using the trained $\mathcal{DNN}$.
\end{enumerate}
The generation of $Y$ via field calculations can take a few days. Afterwards, the trained $\mathcal{DNN}$ 
can predict capacitances very quickly and with sufficient accuracy. This makes it possible to perform a Bayesian exploration of the full parameter space.

Note that this method can be used for detector optimisation in general: not only to fit $\zeta$
but also to find optimal values for other design parameters of a detector.

\section{Detector capacitance matrix}
\label{sec:DetCapMatrix}

\begin{figure}[htbp]
    \centering
    \includegraphics[width=\columnwidth]{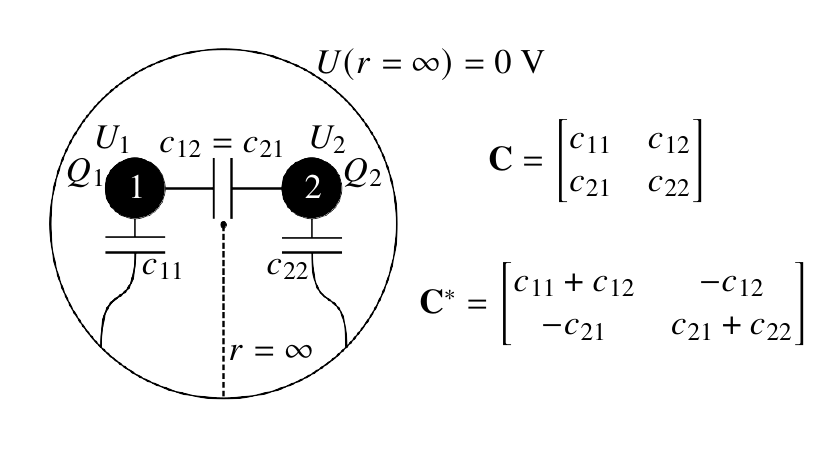}
    \caption{Schematic of the capacitances for a system of two conductors on potentials $U_i$
    with charges $Q_i$ together with the corresponding mutual, $\mathbf{C}$, and Maxwell, $\mathbf{C}^{*}$ capacitance matrices.}
    \label{fig:cap_scheme}
\end{figure}

An HPGe detector with $N \geq 2$ contacts can be seen as a 
system of $N$ conductors which are capacitively coupled. 
A schematic of the capacitances for a system of two conductors is shown in Fig.~\ref{fig:cap_scheme}.

The mutual capacitance, $c_{ij}$, between two conductors $i$ and $j$ is defined as
\begin{linenomath}
\begin{equation}
    c_{ij} = \dfrac{\partial Q_i}{\partial U_j}\,,
    \label{eq:MutualCapacitance}
\end{equation}
\end{linenomath}
where $\partial Q_i$ is the change in charge on conductor $i$ for a change in potential, $\partial U_j$,
of conductor $j$.
The mutual capacitance is symmetric: $c_{ij} = c_{ji}$.
The self capacitance of conductor $i$, $c_{ii}$,
can be understood as a mutual capacitance where the other conductor 
is a grounded closed surface surrounding conductor $i$.
In absence of any surroundings, this surface can be imagined 
as a grounded sphere with an infinite radius.
In case of a typical HPGe detector in a grounded cryostat, the surface is defined through the grounded walls of the cryostat and the grounded parts of the holding structure of the detector.

The different $c_{ij}$ are the elements of the so-called mutual capacitance matrix $\mathbf{C}$.
However, when working with a system of conductors, it is usually not very practical to work with $\mathbf{C}$
as the elements cannot be studied individually.
Therefore, the so-called Maxwell capacitance matrix~\cite{WhitePaperCapMatrix:2020}, 
\begin{linenomath}
\begin{equation}
    \mathbf{C}^{*} = 
    \begin{bmatrix}
    c_{11}^{*} & \dots & c_{1N}^{*} \\
    \vdots & \ddots & \vdots \\
    c_{N1}^{*} & \dots & c_{NN}^{*}
    \end{bmatrix}
    = 
    \begin{bmatrix}
    \sum\limits_{i=1}^{N}c_{1i} & -c_{12} & \dots & -c_{1N} \\
    -c_{21} & \sum\limits_{i=1}^{N}c_{2i} & \dots & -c_{2N}\\
    \vdots & \vdots & \ddots & \vdots \\
    -c_{N1} & -c_{N2} & \dots & \sum\limits_{i=1}^{N}c_{Ni}
    \end{bmatrix}\,,
    \label{eq:MaxwellCapacitanceMatrix}
\end{equation}
\end{linenomath}
is introduced which is generally more practical as it connects the potentials of all conductors, $\vec{U} = (U_1, ..., U_N)$, with 
the charges on all conductors, $\vec{Q} = (Q_1, ..., Q_N)$ via
\begin{linenomath}
\begin{equation}
    \vec{Q} = \mathbf{C}^{*} \cdot \vec{U}
    \label{eq:MaxwellCapacitanceMatrixQVRelation}\,.
\end{equation}
\end{linenomath}

The $\mathbf{C}^{*}$ notation can be distinguished from the $\mathbf{C}$ notation 
through its negative off-diagonal elements with
\begin{linenomath}
\begin{equation}
    c_{ij}^{*} = -c_{ij} \hspace{1cm}\forall\,i \neq j\,.
\end{equation}
\end{linenomath}

The elements of $\mathbf{C}^{*}$ can be calculated~\cite{CapMatrix:2021}
through the weighting potentials\footnote{The weighting potential $\mathcal{W}_i$ of 
contact $i$ in an HPGe detector relates the position of a charge to the induced 
signal in the contact $i$.} of the contacts, $\mathcal{W}_i$:\\
\begin{linenomath}
\begin{equation}
    c_{ij}^{*} = \epsilon_0 \int_{V_\mathrm{W}} \nabla \mathcal{W}_i(\vec{r}) \cdot \epsilon_r(\vec{r}) \cdot \nabla \mathcal{W}_j(\vec{r})\,d\vec{r}\,,
    \label{eq:CalculationOfElementsOfCapMatrix}
\end{equation}
\end{linenomath}
where $\epsilon_0$ is the vacuum permittivity and $\epsilon_r$ the relative permittivity of the medium at position $\vec{r}$. 
The integral is over the closed system volume $V_\mathrm{W}$.

Since the $c_{ij}^{*}$ depend on $\mathcal{W}_i$ and $\epsilon_r(\vec{r})$, they 
depend on the geometry of the detector and its environment.
If the detector is not fully depleted,
the contacts which are touching these undepleted regions become enlarged,
which is an effective change of geometry.
The enlargement of the contacts depends on $\zeta$ and $U_B$.
Thus, for a detector in a fixed environment, the capacitances become 
dependent on two variables:
\begin{linenomath}
\begin{equation}
    c_{ij}^{*} = c_{ij}^{*}(\zeta, U_B)\,.
    \label{eq:CapDependences}
\end{equation}
\end{linenomath}

The absolute values of $c_{ij}^{*}$ decrease with increasing $U_B$.
The $U_B$ at which the detector becomes fully depleted 
is called the full-depletion voltage, $U_B^{\mathrm{fd}}(\zeta)$.
For $U_B > U_B^{\mathrm{fd}}(\zeta)$, the values of $c_{ij}^{*}$ 
basically do not change anymore. 
Thus, there are lower limits on the absolute values for the 
elements of $\mathbf{C}^{*}$ for a given $\zeta$: 
\begin{linenomath}
\begin{equation}
    c_{ij}^{l}(\zeta) = |c_{ij}^{*}(U_B^{\mathrm{fd}}(\zeta))|\,.
    \label{eq:ThresholdCaps}
\end{equation}
\end{linenomath}

It should be noted here that a capacitance is an electrostatic 
quantity and is not frequency dependent.
The reactance is the quantity which introduces a frequency dependence.

\section{The experimental setup K1 and the detector Super-Siegfried}
\label{sec:ExpSetup}

\begin{figure}[htbp]
    \centering
    \includegraphics[width=\columnwidth]{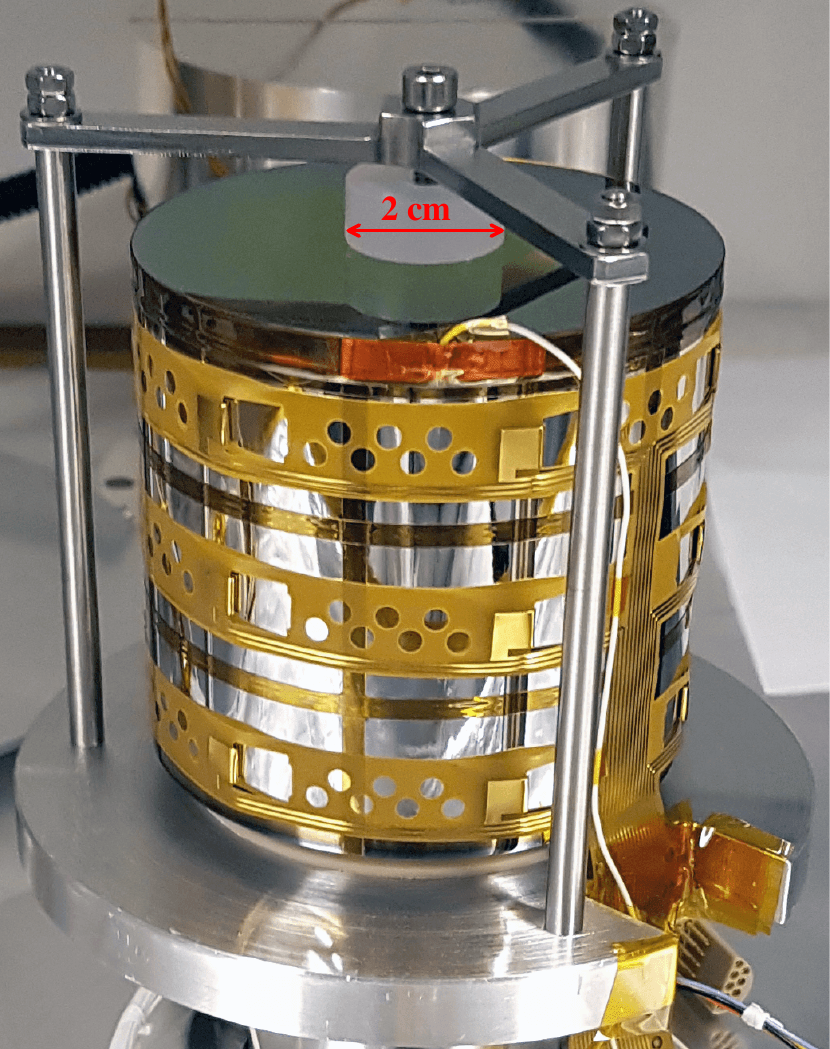}
    \caption{Super-Siegfried within its grounded holding structure as mounted on 
    the grounded base plate.}
    \label{fig:SuSiePicture}
\end{figure}

\begin{figure}[htbp]
    \centering
    \includegraphics[width=\columnwidth]{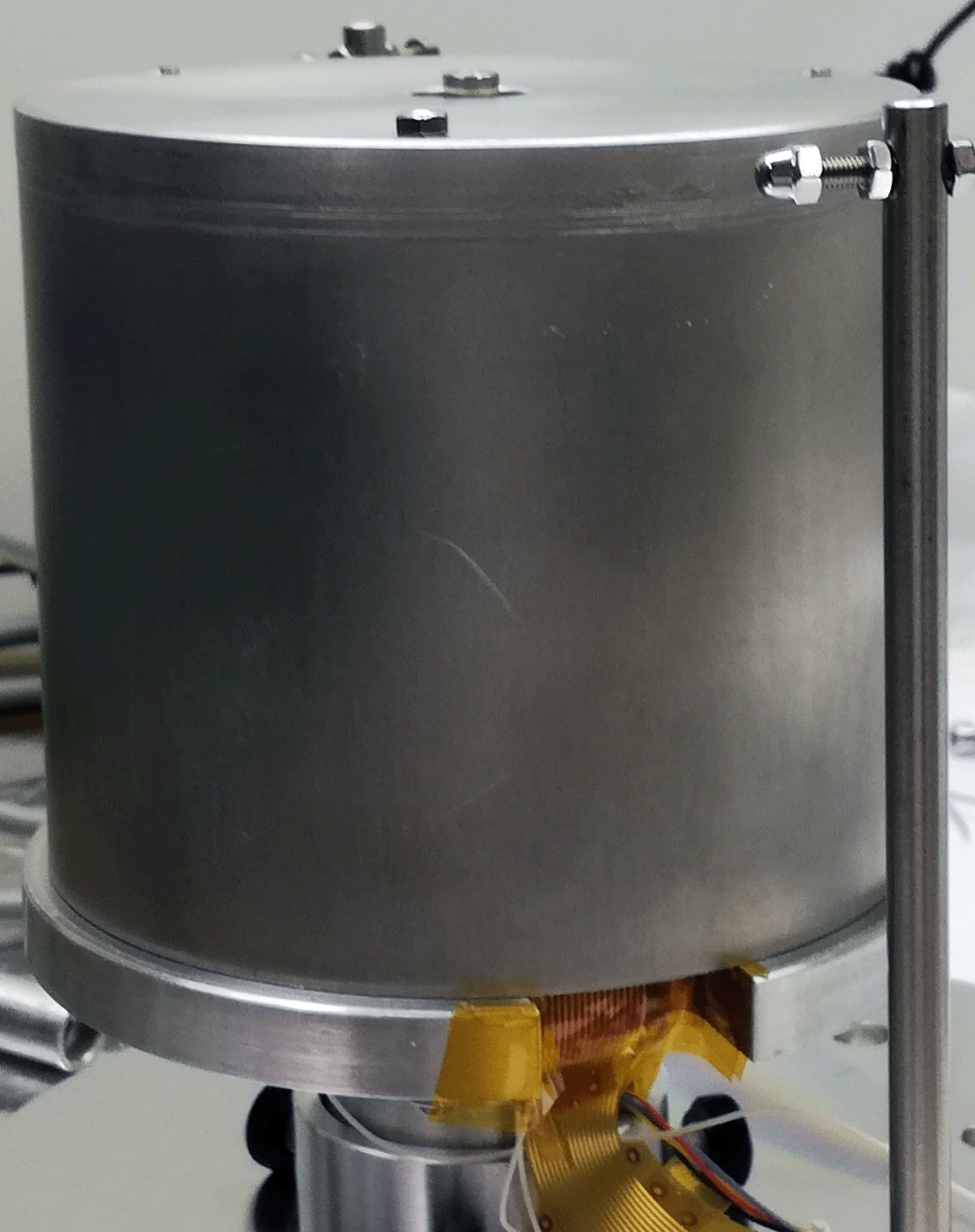}
    \caption{Grounded infrared shield (hat) surrounding Super-Siegfried and
    its holding structure.}
    \label{fig:SuSiePictureUnderHat}
\end{figure}

The test stand K1 is a small vacuum chamber with a cooling finger 
submerged directly in a liquid-nitrogen dewar. 
The $n$-type true-coaxial HPGe detector Super-Siegfried~\cite{Abt:2016trw} was mounted 
on a special base plate which fitted on the cooling finger inside K1.
The detector, together with the necessary holding structure, is depicted in Fig.~\ref{fig:SuSiePicture}.
Closely around the holding structure a so-called hat is placed on top of the base plate,
see in Fig.~\ref{fig:SuSiePictureUnderHat}.
The hat has an inner radius of 55\,mm, an inner height of 105\,mm and 
serves as an infrared shield.
The base plate, the holding structure and the hat are all grounded.

The detector has a length of $l_{\mathrm{D}} = 70$\,mm and a radius of $37.5$\,mm. 
The borehole has a radius of 5\,mm and at the top and the bottom; 
it widens to a radius of 10\,mm within about 3\,mm.
The inner borehole of Super-Siegfried
is the only $n^+$ contact and the mantle is divided into 19 $p^+$ segments.
For the measurements presented in this paper, the segments were not read-out separately,
but were connected together into one single $p^+$ contact. 
The $n^+$ contact is lithium drifted and the $p^+$ segments 
are established through boron implantation.
The manufacturer provided two values for the impurity level 
at the top and at the bottom of the detector:
\begin{linenomath}
\begin{align}
    \zeta_{\mathrm{M}}^{\mathrm{top}} &= 0.44\cdot10^{10} \mathrm{cm}^{-3}\,, \nonumber \\
    \zeta_{\mathrm{M}}^{\mathrm{bot}} &= 1.30\cdot10^{10} \mathrm{cm}^{-3}\,. \nonumber
\end{align}
\end{linenomath}
The operation voltage of the detector suggested by the manufacturer is 3000\,V.

\begin{figure}[htbp]
    \centering
    \includegraphics[width=\columnwidth]{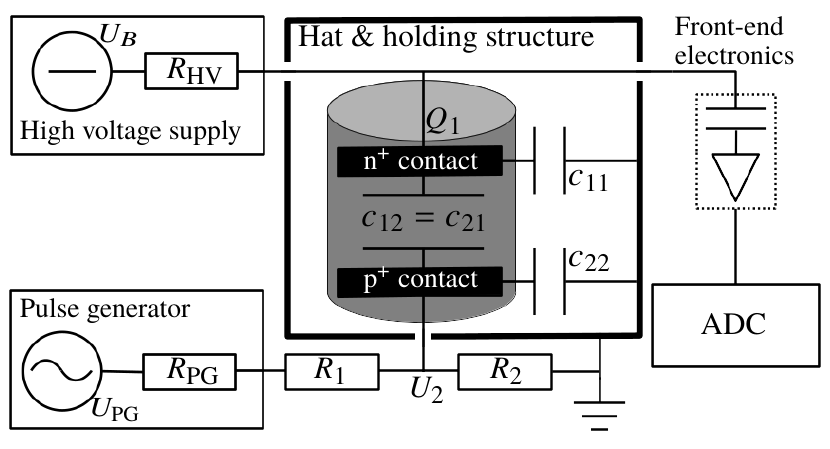}
    \caption{Schematic of the electronics used for the measurements of $c_{12}$ of Super-Siegfried in K1.}
    \label{fig:K1_elec_scheme}
\end{figure}

As the $p^+$ segments are connected to form only one contact, 
the capacitance matrix of the detector is a $2\times2$ matrix as shown in Fig.~\ref{fig:cap_scheme},
where the $n^+$ and the $p^+$ contacts are the two conductors 1 and 2.
The base plate, holding structure and hat form the grounded shell around the two contacts.
This is also shown in Fig.~\ref{fig:K1_elec_scheme}, which shows the
schematic for the measurements of the mutual capacitance between the two contacts, $c_{12}$.
This is similar to what has been described in~\cite{Birkenbach2011HPGEImp}.

$U_B$ is applied to the $n^+$ contact and the $p^+$ contact is held at ground
over a termination resistor, $R_2$.
A pulse generator is connected to the $p^+$ contact over its internal resistor, $R_{\mathrm{PG}}$,
and an additional resistor $R_1$ which serves in combination with $R_2$ as a voltage divider.
To measure $c_{12}$, rectangular pulses with an amplitude of $U_{\mathrm{PG}}$ were generated and
injected into the $p^+$ contact. This corresponds to a change in the potential $U_2$
which translates into a change of charge on the $n^+$ contact, $Q_1$, 
via Eq.~\eqref{eq:MaxwellCapacitanceMatrixQVRelation}:
\begin{linenomath}
\begin{equation}
    Q_1 = - c_{21} \cdot U_2\,.
\end{equation}
\end{linenomath}
Thus, given the electronic circuit, the measured capacitance, $c_{12}^{\textrm{m}}$,
can be calculated for different $U_B$ from the measured $Q_1(U_B)$ as
\begin{linenomath}
\begin{equation}
    c_{12}^{\textrm{m}}(U_B) = \dfrac{-Q_1(U_B)}{U_{2}} = 
        \dfrac{-Q_1(U_B)}{U_{\mathrm{PG}} \cdot \left(\dfrac{R_{2}}{R_{\mathrm{PG}} + R_1 + R_2}\right)}\,,
    \label{eq:measuredC12fromQ}
\end{equation}
\end{linenomath}
where the values for the different components are $R_{\mathrm{PG}} = 50\,\Omega$, $R_{1} = 6190\,\Omega$,
$R_{2} = 51.4\,\Omega\,$ and $U_{\mathrm{PG}} = 126\,\mathrm{mV}$.

$Q_1(U_B)$ was extracted from the induced pulses in the $n^+$ contact as follows:
The $n^+$ contact was connected to a charge-sensitive preamplifier circuit typically used
to read-out germanium detectors.
The amplified signals were recorded with a sampling rate of 250\,MHz and a pulse length of 
20\,\textmu s by a Struck SIS3316~\cite{STRUCK} analog-to-digital converter unit (ADC).
The recorded pulses were inverted ($\Rightarrow  -Q_1$ becomes $+Q_1$ in Eq.~\eqref{eq:measuredC12fromQ}),
corrected for the decay of the charge in the amplification circuit 
and were calibrated~\cite{Hauertmann2017:MasterThesis,Hauertmann2021:PhDThesis}. 
The parameters of the decay correction and calibration of the read-out circuit 
(preamplifier together with the ADC) were determined from pulses of background gamma events 
of known energy of a measurement at $U_B = 3000$\,V. 
An assumption made here is that the parameters of the decay correction and calibration
are independent of $U_B$ or, respectively, $c_{12}$.
After the decay correction and calibration, recorded pulses are often given in units of energy.
For this study, the values were converted into charge pulses, $Q(t)$, in units of charge, pC, 
based on the ionisation energy of germanium.

\begin{figure}[htbp]
    \centering
    \includegraphics[width=\columnwidth]{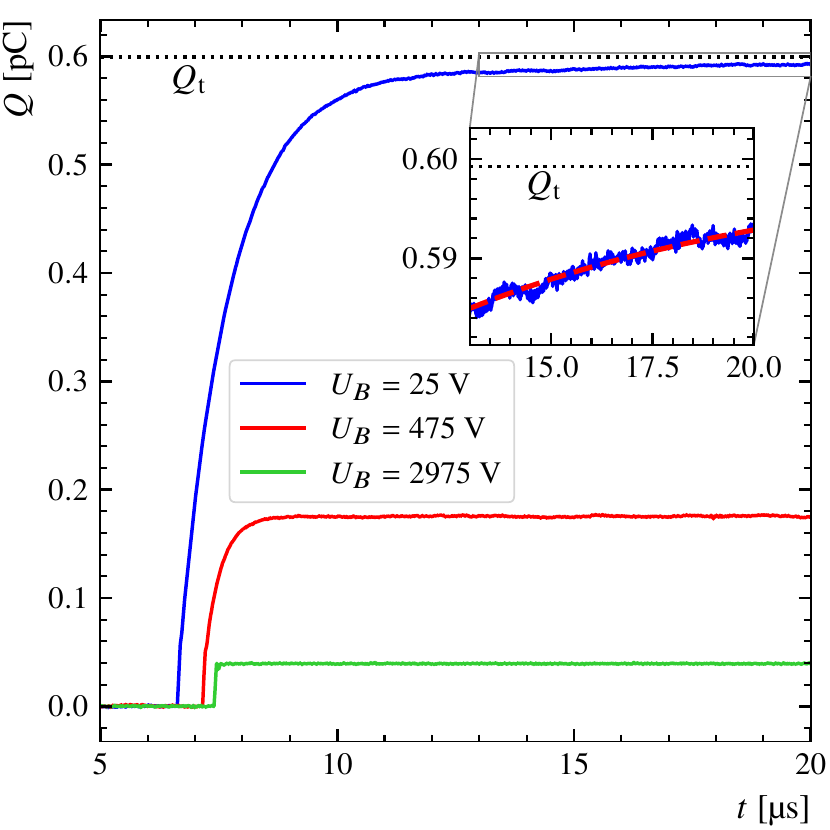}
    \caption{Recorded response of the $n^{+}$ contact of Super-Siegfried in K1 
    to generated rectangular pulses injected into the $p^{+}$ contact for different $U_B$.
    The zoom-in plot shows the fit of $Q(t)$, see Eq.~\eqref{eq:ExtractTotalCharge}, to the tail of the pulse for $U_{B} = 25$\,V.}
    \label{fig:pg_pulse_for_undepl_det}
\end{figure}

For $U_B \geq U_B^{\mathrm{fd}}$, the recorded pulses are rectangular pulses and 
their amplitude corresponds to $Q_1$. 
This is, however, not the case for $U_B < U_B^{\mathrm{fd}}$.
This is shown for three different $U_B$ in Fig.~\ref{fig:pg_pulse_for_undepl_det}.

\begin{figure}[htbp]
    \centering
    \includegraphics[width=\columnwidth]{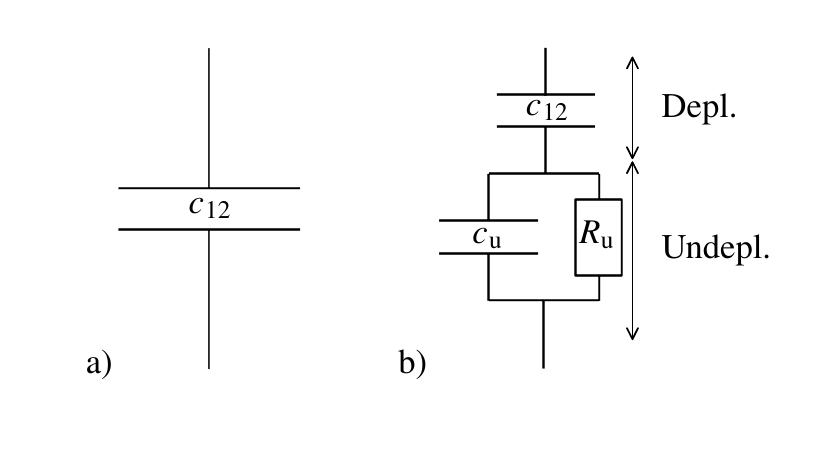}
    \caption{Schematic of the electronics describing a) a fully depleted and b) 
    an only partially depleted detector with only two contacts. 
    The depleted part is modeled with $c_{12}$ while the undepleted part is modeled as 
    an RC circuit with resistance $R_{\mathrm{u}}$ and capacitance $c_{\mathrm{u}}$.}
    \label{fig:undep_det_elec_scheme}
\end{figure}

Partially depleted detectors have to be modeled differently~\cite{RieglerSignalInductionPartialDetectors:2004}
within an electric circuit as shown in Fig.~\ref{fig:undep_det_elec_scheme}.

The undepleted volume can be described as an additional RC component 
introducing also a frequency dependence to the signal and leading to longer pulses.
This can be calculated for one dimensional systems~\cite{RieglerSignalInductionPartialDetectors:2004}.
In reality, that is usually very complicated. However, here, it is not necessary
as we are only interested in the total charge, $Q_t$, flowing through the circuit
which can be determined by fitting the tail of $Q(t)$ with
\begin{linenomath}
\begin{equation}
    Q(t) = Q_{\mathrm{t}} - Q_{\mathrm{u}} \cdot e^{-t/\tau_{\mathrm{u}}}\,,
    \label{eq:ExtractTotalCharge}
\end{equation}
\end{linenomath}
where $Q_t$, $Q_u$ and $\tau_u$ are fit parameters. 
This is also illustrated in Fig.~\ref{fig:pg_pulse_for_undepl_det}.

\section{Measurement of the C-V curve}
\label{sec:MeasuredCVcurve}

\begin{figure}[htbp]
    \centering
    \includegraphics[width=\columnwidth]{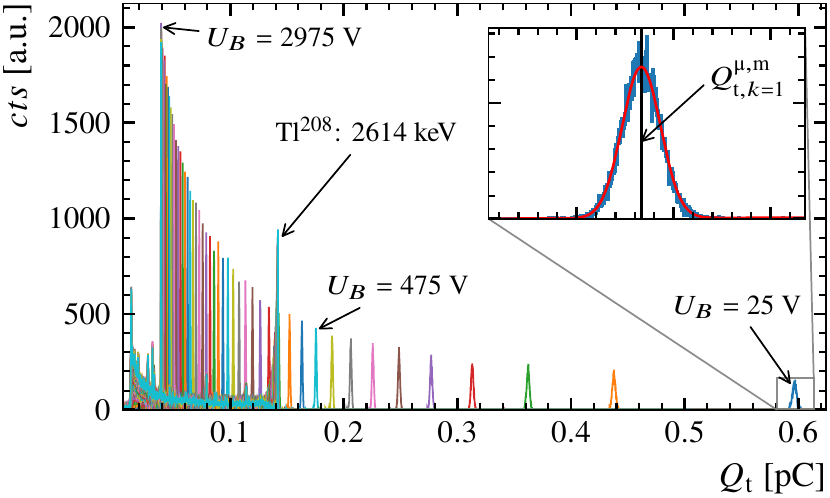}
    \caption{Histograms of $Q_{\mathrm{t}}$ for different $U_B$. 
    The most dominant peak in each spectrum comes from pulser events.
    The $Q_t$ from pulser events decreases for increasing $U_B$.
    For higher bias voltages, also events from background events, mainly from environmental gammas,
    become visible in the range [0, 0.14] pC.
    The peak corresponding to events from 2614\,keV gammas from the Tl$^{208}$ 
    decay is labelled.
    Inset: The fit to the peak of pulser events at $U_B = 25$\,V. 
    The fitted parameter $Q_{\mathrm{t},k=1}^{\upmu,\mathrm{m}}$ is shown as a vertical line.}
    \label{fig:pg_pulse_amp_hists_and_fit}
\end{figure}

The detector was operated at 60 different bias voltages,
\begin{linenomath}
\begin{equation}
    U_{B,k} = 25\,\mathrm{V} + (k-1) \cdot 50\,\mathrm{V} \hspace{0.5cm} \forall~k \in \{1, 2, ..., 60\}\,.
    \label{eq:SetOfBiasVoltages}
\end{equation}
\end{linenomath}
Each measurement lasted 600\,s. 
For lower $U_{B,k}$, most of the observed pulses were induced by the pulse generator.
For increasing $U_{B,k}$, more pulses induced by gammas from natural radioactivity were recorded 
as the depleted volume and, thus, the active volume of the detector increased.
However, the peak created by the pulse generator was always clearly identifiable.
The spectra of $Q_t$ of all pulses from all 60 measurements are shown in Fig.~\ref{fig:pg_pulse_amp_hists_and_fit}.
For all measurements, the events induced by the pulse generator form a peak. 
The mean values of these peaks, $Q_{\mathrm{t},k}^{\upmu,\mathrm{m}}$, 
were determined by fits of scaled normal distributions as shown in Fig.~\ref{fig:pg_pulse_amp_hists_and_fit}.

\begin{figure}[htbp]
    \centering
    \includegraphics[width=\columnwidth]{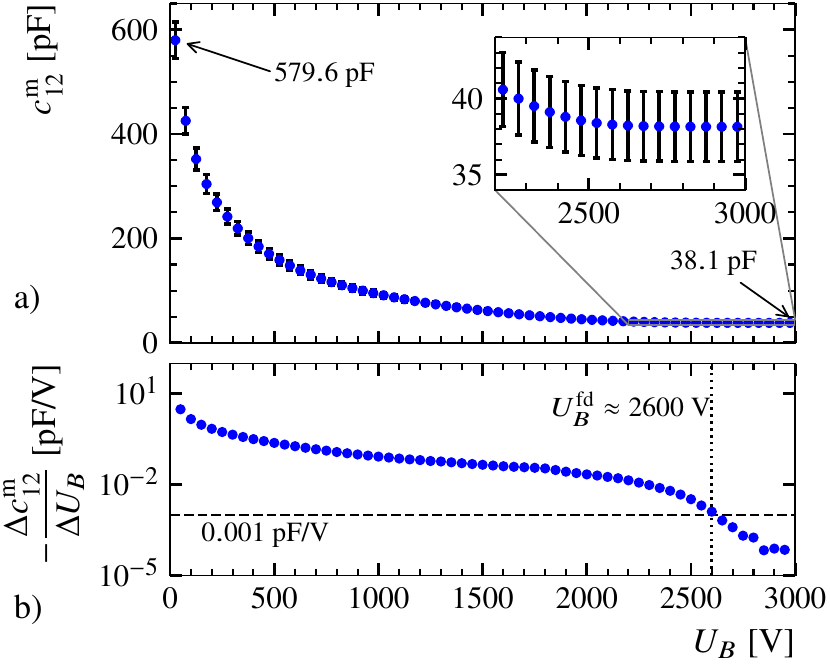}
    \caption{a) C-V curve of Super-Siegfried as measured in K1. 
    The error bars represent uncertainties conservatively determined by assuming 3\% uncertainties on $Q_{\mathrm{t},k}^{\upmu,\mathrm{m}}$,
    $R_{\mathrm{PG}}$, $R_{1}$, $R_{2}$ and $U_{\mathrm{PG}}$ and using Gaussian error propagation.
    The uncertainties are highly correlated. b) Negative differences, $- \Delta c_{12}^{\mathrm{m}}/\Delta U_B$, between the data points in a).}
    \label{fig:dataCVScan}
\end{figure}

Using Eq.~\eqref{eq:measuredC12fromQ}, the determined $Q_{\mathrm{t},k}^{\upmu,\mathrm{m}}$
were converted to $c_{12}^{\mathrm{m,k}}$ for all $U_{B,k}$.
The resulting C-V curve, $c_{12}^{\mathrm{m},k}$, is shown in Fig.~\ref{fig:dataCVScan}a.

As mentioned earlier, in theory, $c_{12}$ should not change anymore for $U_B > U_B^{\mathrm{fd}}$. 
This does not take into account that the contacts are regions of the detector, which are
doped more than three orders of magnitude higher than the bulk. 
With increasing $U_B$, also very small volumes of the contacts become depleted. 
However, for reasonable values of $U_B$ below the break-through voltage, the contacts
never become completely depleted and $-\partial c_{12}^{\mathrm{m}} / \partial U_B$ never 
becomes entirely zero. This is demonstrated in Fig.~\ref{fig:dataCVScan}b.
Therefore, it is not trivial in general to define $U_B^{\mathrm{fd}}$. 
Here, we define it as $-\Delta c_{12}^{\mathrm{m}} / \Delta U_B \overset{!}{=} 10^{-3}$\,pF/V: $U_B^{\mathrm{fd}}\approx 2600$\,V. 
For $U_B > 2600$\,V, simulations are expected to result in a fully depleted detector for the correct $\zeta$. 

\section{Simulation of the C-V curve}
\label{sec:DeplHandlingInSSD}

The C-V curve for a given $\zeta$ was simulated with \emph{SolidStateDetectors.jl} (SSD).
Since version v0.7, SSD can be used to calculate $\mathbf{C}^{*}$ of a detector 
while taking the influence of the environment into account. 
Since v0.8, it can also perform the required 3d field calculations for $\Phi$ and $\mathcal{W}_i$ efficiently on GPUs.

In SSD, $\Phi$ and $\mathcal{W}_i$ are calculated by solving Gauss's law on adaptive 3d (cylindrical or Cartesian)
grids via the iterative successive over-relaxation (SOR) algorithm~\cite{Hauertmann2021:PhDThesis}.
As the potentials are calculated on grids, the integral in Eq.~\eqref{eq:CalculationOfElementsOfCapMatrix} 
is turned into a sum over the grid of $\mathcal{W}_i$. The gradient of $\mathcal{W}_j$ 
is determined through interpolation onto the grid of $\mathcal{W}_i$ 
as the two final grids are usually not identical 
due to the adaptive grid refinement.

In order to calculate the elements of the matrix via Eq.~\eqref{eq:CalculationOfElementsOfCapMatrix},
$\Phi$ and all $\mathcal{W}_i$ have to be calculated first.
In contrast to $\mathcal{W}_i$, $\Phi$ does not occur in Eq.~\eqref{eq:CalculationOfElementsOfCapMatrix}. 
It is, however, required to determine the depleted volume which depends on $\zeta$ and $U_B$~\cite{Hauertmann2021:PhDThesis}.
This information is then passed to the calculation of the weighting potentials.
In the calculation of the weighting potentials, the relative permittivity, $\epsilon_r$, 
inside undepleted volumes was scaled by $10^5$ as an approximation of infinity. 
This makes these areas quasi-conductive and results in equal potentials over these volumes.
Thus, undepleted volumes in touch with a contact become extensions of this contact and 
will be on the same potential as applied to the contact.

\begin{figure}[htbp]
    \centering
    \includegraphics[width=\columnwidth]{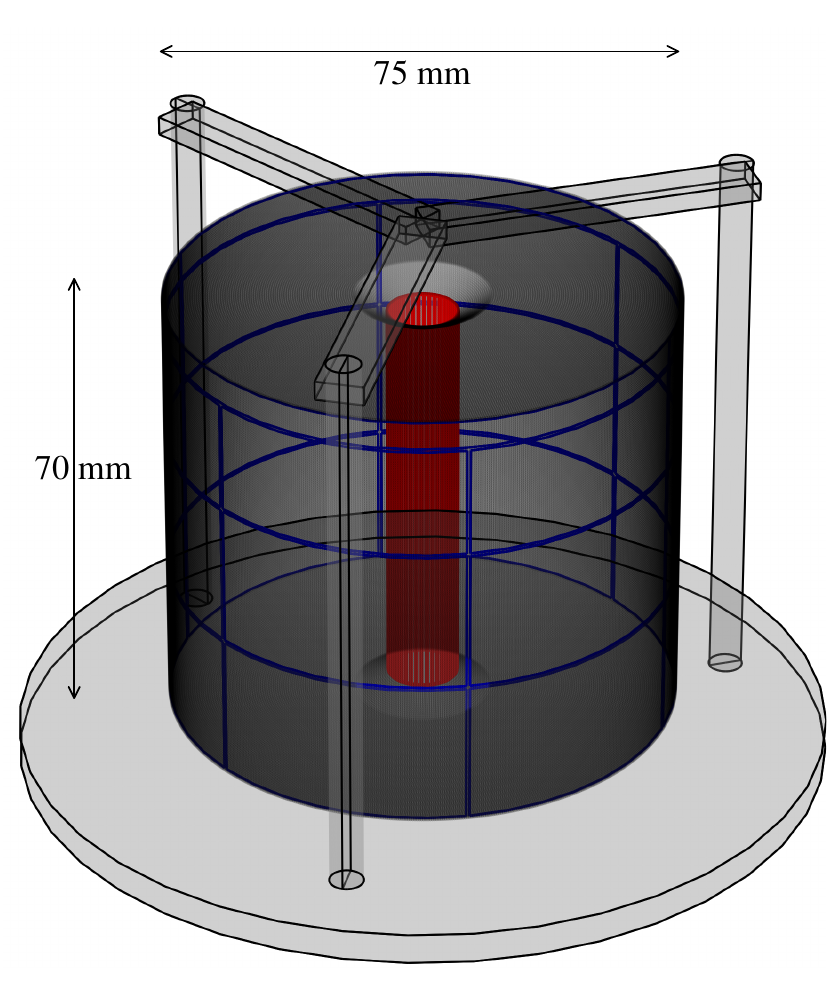}
    \caption{The geometry of the HPGe detector Super-Siegfried (dark grey), 
        together with its holding structure and base plate (light grey) 
        as implemented in SSD. The blue lines are segment boundaries of the $p^+$ contact and the red cylinder is the $n^+$ contact.}
    \label{fig:SuSie_in_K1_SSD}
\end{figure}

The detector Super-Siegfried and the grounded holding structure and base plate 
as implemented in SSD are shown in Fig.~\ref{fig:SuSie_in_K1_SSD}.
For the simulations, a cylindrical grid was chosen and the grid was limited to the dimensions of the hat. 
Fixed boundary conditions of 0\,V were set at the outer edge in $r$ and at both edges in $z$ 
to mimic the grounded closed shell of the hat and base plate.
In $\varphi$, the grid was limited from $0^{\circ}$ to $120^{\circ}$ and periodic boundary conditions were set.

It is very important to define the geometry of the detector, especially the geometry of the contacts, 
as realistically possible as it also influences the capacitance. 
Particularly, $c_{12}^{l}$ depends on the exact geometry. 
For all simulations presented in this paper, the $p^+$ contacts of the detector were fixed 
to a thickness of $0.5$\,\textmu m as measured previously~\cite{Hauertmann2021:PhDThesis}.
Lithium drifted contacts typically have thicknesses of $\mathcal{O}(\mathrm{mm})$. 
This thickness was not measured previously since irradiation of the inner borehole is not simple to achieve.
Since the $n^+$ contact thickness, $d_{\mathrm{Li}}$, is on the mm scale, 
it impacts $c_{12}^{l}$ on a measurable scale.
The $n^+$ contact geometry is implemented as a tube with the inner radius being fixed at the borehole of the detector.
The thickness of the tube, $d_{\mathrm{Li}}$, is a free parameter in the fit presented in this paper.
The contact does not cover the widening of the borehole. 
At the bottom and the top of the contact, the outer edge of the tube is rounded off.

\begin{figure}[htbp]
    \centering
    \includegraphics[width=\columnwidth]{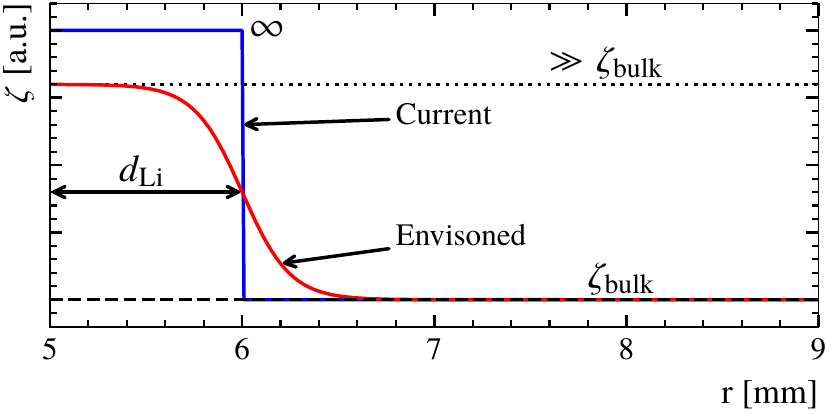}
    \caption{One-dimensional illustration of the envisioned modelling of the $n^+$ layer for future simulations 
    with SSD via a continuous increase of the impurity density from the impurity density of the bulk, $\zeta_{\mathrm{bulk}}$,
    to the impurity density of the $n^+$ contact.
    The hard transition corresponds to the current implementation of the $n^+$ layer in SSD,
    where the potential values inside the contact volume are fixed to the set contact potential.}
    \label{fig:li_contact_modeling}
\end{figure}

In the field calculation, the potential values of grid points inside 
the defined volumes of the contacts are fixed to the potential applied to the corresponding contact. 
In principle, the $n^+$ contact could instead be modeled through $\zeta$.
The current implementation of a fixed potential inside the volume corresponds to a jump from the bulk impurity 
density to infinity at the surface of the contact volume as shown in Fig.~\ref{fig:li_contact_modeling}.
In future, it is envisioned to smooth this hard edge transition by adding some continuous function
to the impurity at the $n^+$ contact\footnote{The $p^+$ contact thickness is on a much smaller scale, $\mathcal{O}($\textmu m$)$,
and, thus, it would not be feasible to resolve a smooth edge transition on that scale.}. 
This smooth edge transition is also shown in Fig.~\ref{fig:li_contact_modeling}.
However, this is not yet part of the studies presented in this paper.

In SSD, custom signed impurity-densities can be defined where the sign of the given density determines the sign of the fixed space 
charges of the minority charge carriers at the specific location. Thus, the sign is used to specify 
the type ($n$ or $p$) of the semiconductor at a specific location, $\vec{r} = (r, \varphi, z)$:
\begin{linenomath}
\begin{align}
    \zeta(\vec{r}) > 0\,\, \mathrm{cm}^{-3} \Leftrightarrow ~&n\mathrm{-type\,region}\,,\label{eq:sign-ntype}\\
    \zeta(\vec{r}) < 0\,\, \mathrm{cm}^{-3} \Leftrightarrow ~&p\mathrm{-type\,region}\,.\label{eq:sign-ptype}
\end{align}
\end{linenomath}

Typically, in simulations of HPGe detectors, a simple linear or quadratic change of $\zeta$
is assumed along the crystal pulling axis $z$,
based on certain levels of impurity provided by the manufacturer for some $z$ values. 
A radial component is usually not assumed.
Thus, for the detector Super-Siegfried, the signed impurity-density model based on 
manufacturer values becomes
\begin{linenomath}
\begin{equation}
    \zeta_{\mathrm{M}}(z) = \zeta_{\mathrm{M}}^{\mathrm{bot}}(z) + \dfrac{\zeta_{\mathrm{M}}^{\mathrm{top}}(z) - \zeta_{\mathrm{M}}^{\mathrm{bot}}(z)}{l_{\mathrm{D}}} \cdot z\,,
    \label{eq:ManufacturerImpurityDensity}
\end{equation}
\end{linenomath}
which comprises a linear profile in $z$ and no modulation in $r$.

\begin{figure}[htbp]
    \centering
    \includegraphics[width=\columnwidth]{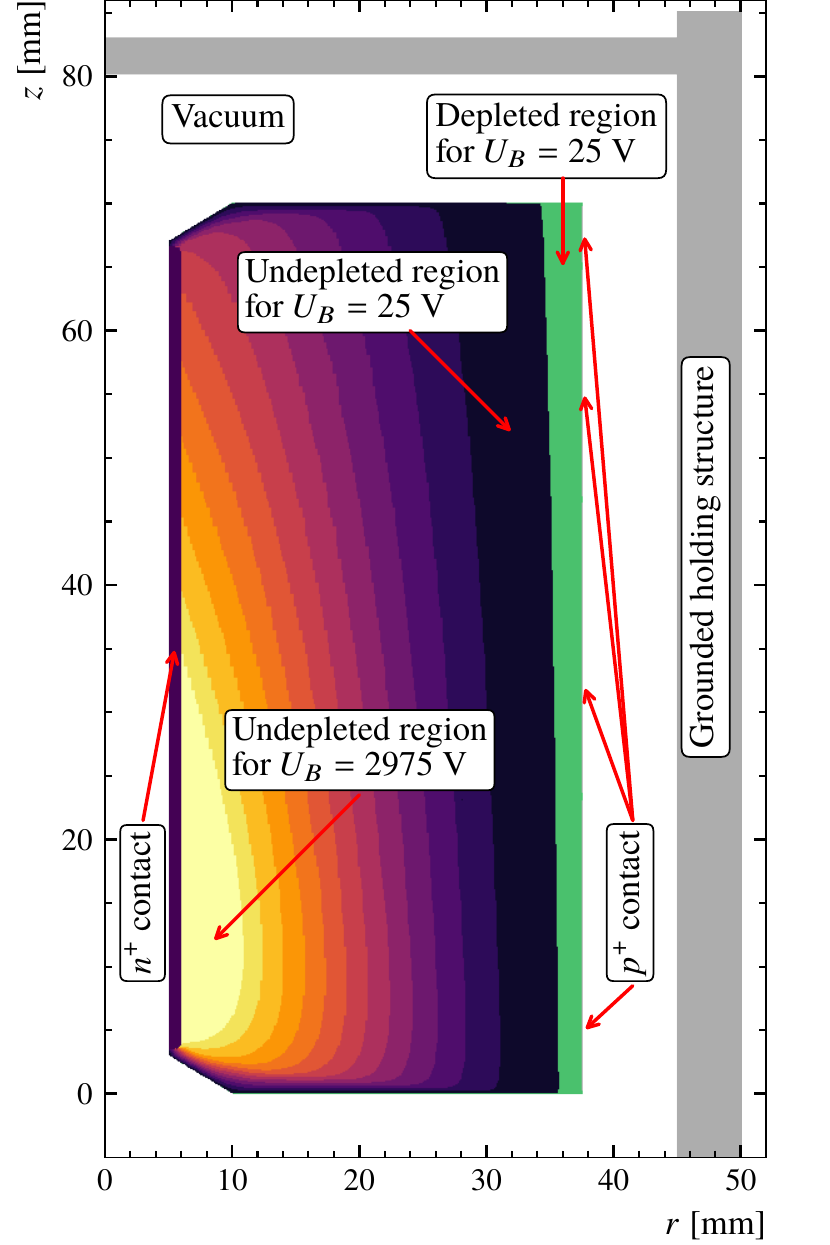}
    \caption{Cross-section of Super-Siegfried mounted in K1 at $\varphi = 36.7^{\circ}$
        showing the undepleted volumes for $\zeta_{\mathrm{M}}$ and selected $U_{B,k}$ 
        in steps of 250\,V (200\,V for the last step to 2975\,V) as differently shaded areas.
    }
    \label{fig:ssd_impM_point_types}
\end{figure}

The depleted volumes for $\zeta_{\mathrm{M}}$ and $d_{\mathrm{Li}} = 1$\,mm 
were calculated with SSD for all $U_{B,k}$.
The undepleted volumes are shown in Fig.~\ref{fig:ssd_impM_point_types} for selected $U_{B,k}$.

It shows how the detector depletes from the mantle of the detector towards the borehole.
It also shows that even for a bias voltage of $2975$\,V the detector does not become 
fully depleted.
Thus, the overall impurity level of $\zeta_{\mathrm{M}}$ seems to be too high 
since $U_{B}^{\mathrm{fd}}$ was determined to be $\approx 2600$\,V.

\begin{figure}[htbp]
    \centering
    \includegraphics[width=\columnwidth]{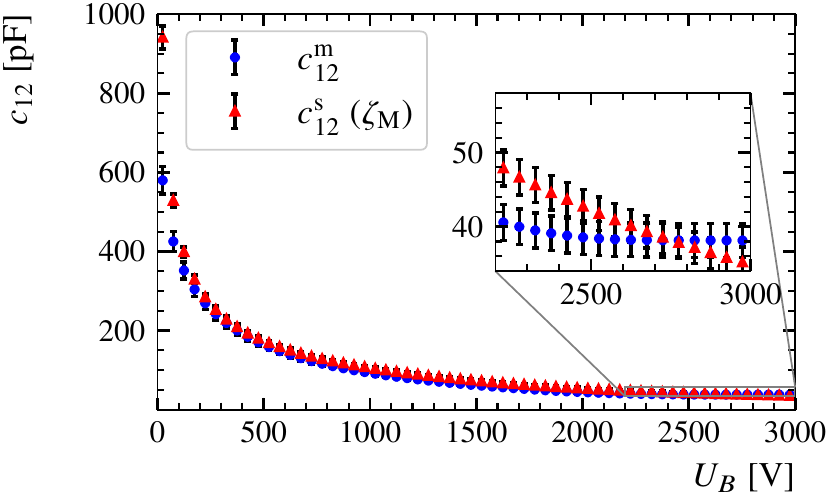}
    \caption{C-V curve, $c_{12}^{\mathrm{s},k}$, as simulated with SSD for $\zeta_{\mathrm{M}}$ together 
    with the measured C-V curve, $c_{12}^{\mathrm{m},k}$, already shown in Fig.~\ref{fig:dataCVScan}a.}
    \label{fig:data_vs_SSD_manufacturer_CV_scan}
\end{figure}

The simulated C-V curve, $c_{12}^{\mathrm{s},k}$, in comparison to the measured C-V curve, 
is shown in Fig.~\ref{fig:data_vs_SSD_manufacturer_CV_scan}.
An uncertainty of 
\begin{linenomath}
\begin{equation}
    \sigma_c^{\mathrm{s}}(c_{12}^{\mathrm{s}}) = c_{12}^{\mathrm{s}} \cdot 1\% + 1\,\mathrm{pF}
    \label{eq:CapUncertaintySSD}    
\end{equation}
\end{linenomath}
is assigned to $c_{12}^{\mathrm{s},k}$. The absolute uncertainty of $1$\,pF is motivated by a possible imperfect 
implementation of the geometry in comparison to reality. The 1\% relative uncertainty is 
motivated by studies on the simulated capacitance for different levels of the fineness of the 
final grids of the calculated fields.
A relative uncertainty is chosen for this source of uncertainty because the depleted volume is smaller 
for lower bias voltages (larger capacitances) requiring a finer refinement.
In SSD, the refinement of the grids in the field calculations can be tuned~\cite{SSD:Website}. 
The following settings were used for the simulations presented in this paper:
\begin{linenomath}
\begin{align*}
    convergence\_limit &= 10^{-7}\,,\\
    refinement\_limits &= [0.2, 0.2, 0.1, 0.1, 0.1,\\
    &\hspace{0.53cm}0.05, 0.03, 0.02, 0.01]\,,\\
    max\_distance\_ratio &= 3.0\,, \\  
    min\_tick\_distance &= (10\,{\hbox{\textmu}}\mathrm{m}, 1^{\circ}, 10\,{\hbox{\textmu}}\mathrm{m})\,.
\end{align*}
\end{linenomath}

Figure~\ref{fig:data_vs_SSD_manufacturer_CV_scan} shows that the simulation predicts that 
the detector is not yet fully depleted for any $U_{B,k}$
as the simulated C-V curve still decreases and does not reach its lower limit at $U_{B,60}$.
However, at $U_{B,60}$, the simulated capacitance is lower than 
the measured capacitance limit $c_{12}^{l,m} = c_{12}^{\mathrm{m},60}$. 
This indicates that $d_{\mathrm{Li}}$ might be larger or that the implemented geometry 
does not perfectly describe reality.
For low bias voltages, $c_{12}^{\mathrm{s},k}$ is larger than $c_{12}^{\mathrm{m},k}$.
This means that the detector depletes faster in reality than in the simulation,
indicating that the $\zeta_{\mathrm{M}}$ is too large at larger radii.

\section{Model function for the impurity density}
\label{sec:ImpurityDensityModel}

The comparison between $c_{12}^{\mathrm{m},k}$ and $c_{12}^{\mathrm{s},k}$ for $\zeta_{\mathrm{M}}$ 
suggests that the impurity density is not that high and not constant in $r$ and a more complex model for $\zeta$ is required.

\begin{figure}[htbp]
    \centering
    \includegraphics[width=\columnwidth]{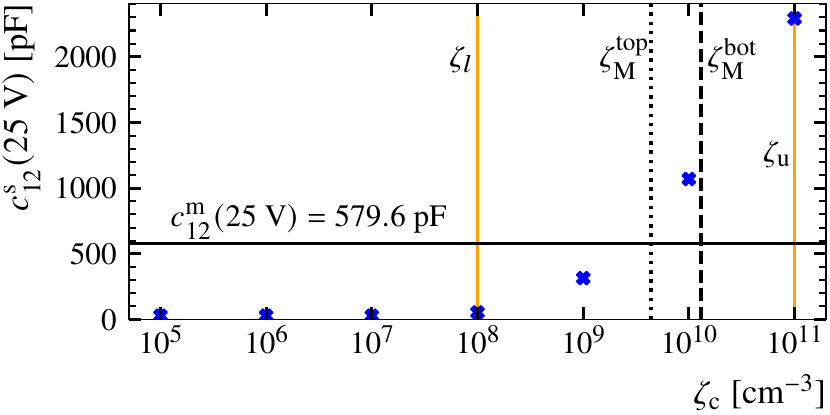}
    \caption{Capacitances $c_{12}^{\mathrm{s}}$ for ($U_B = 25$\,V, $d_{\mathrm{Li}} = 1\,$mm) 
        and for different constant levels of $\zeta_{\mathrm{c}}$ as simulated with SSD. 
        The vertical solid lines mark the parameter space of the impurity level where $c_{12}$ is sensitive to $\zeta$. The impurity levels $\zeta_{\mathrm{M}}^{\mathrm{bot}}$ and 
        $\zeta_{\mathrm{M}}^{\mathrm{top}}$ provided by the manufacturer are also indicated.}
    \label{fig:impurity_level_sensitivity}
\end{figure}

Assuming a constant impurity density $\zeta_{\mathrm{c}}$, the range of sensitivity of the
voltage scan to the absolute impurity levels can be estimated. 
For this, $c_{12}^{\mathrm{s}}$ is calculated for ($U_B = 25$\,V, $d_{\mathrm{Li}} = 1\,$mm) and for different $\zeta_{\mathrm{c}}$ as shown in Fig.~\ref{fig:impurity_level_sensitivity}.

For $d_{\mathrm{Li}} = 1$\,mm, the simulated limit capacitance, $c_{12}^{l,s}$, can be calculated 
with $\zeta_{\mathrm{c}} = 0\,\mathrm{cm}^{-3}$.
Even for $\zeta_{\mathrm{c}} = 10^{8}\,\mathrm{cm}^{-3}$, ${c_{12}^{\mathrm{s}} \approx c_{12}^{l,s}}$
meaning that the detector is already fully depleted at ${U_B = 25}$~V. 
Thus, this level can be seen as a lower limit, $\zeta_{l}$, for the parameter space of impurity levels, 
\begin{linenomath}
\begin{equation}
    \zeta_{l} \coloneqq 10^{8}\,\mathrm{cm}^{-3}\,.    
    \label{eq:ImpLowerLimit}
\end{equation}
\end{linenomath}
At $\zeta_{\mathrm{c}} = 10^{11}\,\mathrm{cm}^{-3}$, $c_{12}^{\mathrm{s}}$ is much larger 
than the measured capacitance at $U_B = 25$\,V. 
Thus, this level can be seen as an upper limit, $\zeta_{\mathrm{u}}$, for the parameter space of impurity levels, 
\begin{linenomath}
\begin{equation}
    \zeta_{\mathrm{u}} \coloneqq 10^{11}\,\mathrm{cm}^{-3}\,.    
    \label{eq:ImpUpperLimit}
\end{equation}
\end{linenomath}
Figure~\ref{fig:impurity_level_sensitivity} also shows that a logarithmic change in $\zeta_{\mathrm{c}}$ between $\zeta_{l}$ and $\zeta_{\mathrm{u}}$ results in a change of $c_{12}^{\mathrm{s}}$ on a linear scale.
Therefore, the ($r,z$)-dependent signed impurity-density, 
$\zeta_{\mathrm{RZ}}(r, z|s)$, is defined as
\begin{linenomath}
\begin{equation}
    \zeta_{\mathrm{RZ}}(r, z|g) = \tanh(g\cdot x(r, z))\cdot 10^{|x(r, z)|}\,,
    \label{eq:ImpDensityModel}
\end{equation}
\end{linenomath}
where $g$ is set to 1000 for this study and $x(r, z)$ is a function to and above $-\zeta_{l}$.
model $\zeta_{\mathrm{RZ}}$ on a logarithmic scale.
The sign of $x(r, z)$ is used to determine the type of the semiconductor at $(r, z)$, see Eq.~\eqref{eq:sign-ntype}
and Eq.~\eqref{eq:sign-ptype}.

\begin{figure}[htbp]
    \centering
    \includegraphics[width=\columnwidth]{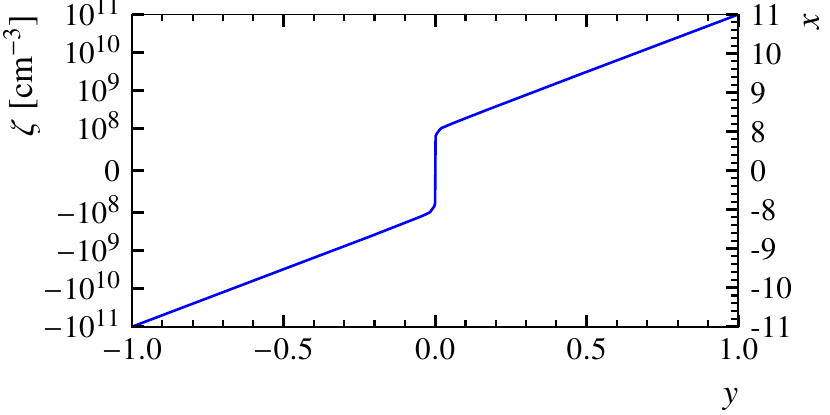}
    \caption{
        Signed impurity-density $\zeta_{\mathrm{RZ}}(\mathcal{T}(y))$, see Eq.~\eqref{eq:ImpDensityModel}, on the left axis
        and the parameter transformation $x = \mathcal{T}(y)$, see Eq.~\eqref{eq:ParameterTrafoFunction}, on the right axis
        as a function of $y$.}
    \label{fig:parameter_transformation_uniform_to_log}
\end{figure}

According to the simulation, the measurement is not sensitive to impurity 
densities below $\zeta_{l}$ and above $-\zeta_{l}$\footnote{The 
same sensitivity is assumed for $p$-type densities as for the inverted system, 
inverted charge distribution and contact potentials, the same values for the capacitances will be calculated.}.
A parameter transformation, $\mathcal{T}$, is defined to model $\zeta_{\mathrm{RZ}}$ between $[-\zeta_{\mathrm{u}},\zeta_{\mathrm{u}}]$ and significantly reduce the influence of the parameter interval
$[-\zeta_{l},\zeta_{l}]$:
\begin{linenomath}
\begin{align}
    &x(r, z) = \mathcal{T}(y(r, z))\,,\label{eq:ParameterTrafo}\\
    &\mathcal{T}(y|v,x_{l}, x_{\mathrm{u}}) = \sign(y) \cdot x_{l} \cdot |y|^{\frac{1}{v}} + (x_{\mathrm{u}} - x_{l}) \cdot y\,,\label{eq:ParameterTrafoFunction}
\end{align}
\end{linenomath}
with $x_{l} = \log_{10}(\zeta_{l}\cdot\mathrm{cm}^3)\hspace{0.05cm}= 8$,
${x_{\mathrm{u}}=\log_{10}(\zeta_{\mathrm{u}}\cdot\mathrm{cm}^{3})=11}$, ${v = 1000}$ and
$y(r, z)$ is a function to model $\zeta_{\mathrm{RZ}}$ on the linear interval [$-1,1$].
The transformation is shown in Fig.~\ref{fig:parameter_transformation_uniform_to_log}.
A linear change in $y$ will result in a logarithmic change in $\zeta_{\mathrm{RZ}}$ which 
causes a change in $c_{12}$ on a linear scale.

The spatial dependence of the model $\zeta_{\mathrm{RZ}}(r,z)$ is implemented as a spatial
dependence of $y$:
\begin{linenomath}
\begin{equation}
    y(r, z) = y^{\mathrm{bot}}(r) + \dfrac{y^{\mathrm{top}}(r) - y^{\mathrm{bot}}(r)}{l_{\mathrm{D}}} \cdot z\,
    \label{eq:Y_RZ}
\end{equation}
\end{linenomath}
where the $r$ dependence is modeled by $y^{\mathrm{bot}}(r)$ at $z = 0\,$mm and
$y^{\mathrm{top}}(r)$ at $z = l_{\mathrm{D}}$. Both, $y^{\mathrm{bot}}(r)$ and $y^{\mathrm{top}}(r)$, are 
modeled as two cubic splines defined for four specific radial positions $r_{\mathrm{b,1}} = 20$\,mm, 
$r_{\mathrm{b,2}} = 28$\,mm, $r_{\mathrm{b,3}} = 33$\,mm and $r_{\mathrm{b,4}} = 37.5$\,mm.
The gradient of the splines at their left boundary, $r_{\mathrm{b,1}}$, is set to zero,
the gradient at the right boundary, $r=37.5$\,mm, is not fixed.

Thus, the model is defined by a set of 8 parameters, $p_{\zeta}$, 
either defined in impurity levels
\begin{linenomath}
\begin{equation}
     p_{\zeta} = \left(\zeta_1^{\mathrm{bot}}, \zeta_2^{\mathrm{bot}}, \zeta_3^{\mathrm{bot}}, \zeta_4^{\mathrm{bot}},
           \zeta_1^{\mathrm{top}}, \zeta_2^{\mathrm{top}}, \zeta_3^{\mathrm{top}}, \zeta_4^{\mathrm{top}}\right)\,,
\end{equation}
\end{linenomath}
or in values of $y$,
\begin{linenomath}
\begin{equation}
    p_{\zeta_y} = \left(y_1^{\mathrm{bot}}, y_2^{\mathrm{bot}}, y_3^{\mathrm{bot}}, y_4^{\mathrm{bot}},
           y_1^{\mathrm{top}}, y_2^{\mathrm{top}}, y_3^{\mathrm{top}}, y_4^{\mathrm{top}}\right)\,,
\end{equation}
\end{linenomath}
as they can be transformed into each other via Eq.~\eqref{eq:ImpDensityModel}
to Eq.~\eqref{eq:ParameterTrafoFunction} and their inverse functions.

\begin{figure}[htbp]
    \centering
    \includegraphics[width=\columnwidth]{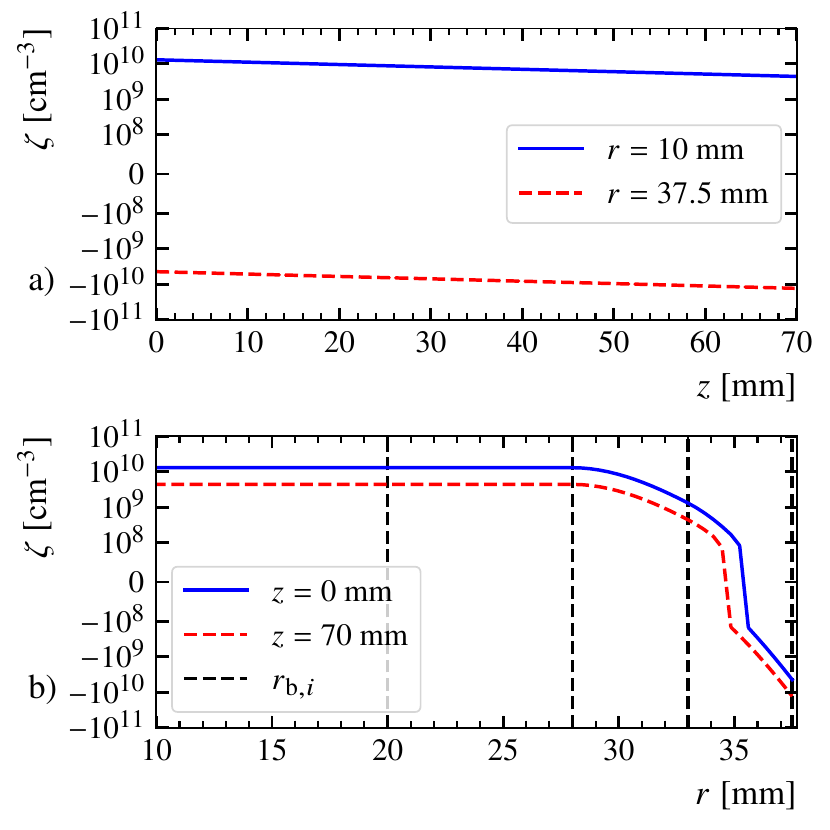}
    \caption{Example impurity distribution $\zeta_{\mathrm{RZ}}$ for $p_{\zeta,\mathrm{E}}$
        as a function of a)~$z$ and b)~$r$. See text for details.}
    \label{fig:example_impurity_density}
\end{figure}

An example impurity distribution of $\zeta_{\mathrm{RZ}}$ based
on the values provided by the manufacturer is 
shown in Fig.~\ref{fig:example_impurity_density} with 
\begin{linenomath}
\begin{equation*}
p_{\zeta,\mathrm{E}} = \left(\zeta_{\mathrm{M}}^{\mathrm{bot}}, \zeta_{\mathrm{M}}^{\mathrm{bot}}, 
                  \zeta_{\mathrm{M}}^{\mathrm{bot}}/10, -\zeta_{\mathrm{M}}^{\mathrm{top}},
                  \zeta_{\mathrm{M}}^{\mathrm{top}}, \zeta_{\mathrm{M}}^{\mathrm{top}}, 
                  \zeta_{\mathrm{M}}^{\mathrm{top}}/10, -\zeta_{\mathrm{M}}^{\mathrm{bot}}\right)\,.
      \label{eq:ExampleImpurityProfileParameters}
\end{equation*}
\end{linenomath}

The model allows to modulate the bulk impurity density of a detector including a possible 
boundary between $n$-type and $p$-type volumes 
as demonstrated with the example density, see Fig.~\ref{fig:example_impurity_density}.
There, the main bulk of the detector is $n$-type but is $p$-type close to the mantle.
For this detector, such a $p$-type volume close to the mantle is motivated by two aspects. 
First, when pulling the crystal via the Czochralski method, there could be some radial modulation of the impurities due to the process.
Especially, since natural germanium is $p$-type and $n$-type 
dopants have to be added to the molten germanium.
Secondly, the $p^+$ contacts are heavily over-doped layers, $\mathcal{O}(10^{12}\,\mathrm{cm}^{-3})$ or even higher.
The thickness of the undepleted boron layers is very small, $0.5$\,\textmu m, 
but there could be diffusion.
This could lead to impurities reaching the magnitude of the bulk densities
, $\mathcal{O}(10^{8-10}\,\mathrm{cm}^{-3})$,
penetrating deeper, $\mathcal{O}(\textrm{mm})$, 
into the $n$-type germanium leading to compensation and type conversion.

\section{Deep neural network for fast capacitance predictions}

For a given set of values for ($d_{\mathrm{Li}}$, $p_{\zeta_y}$),
a C-V curve can be calculated with SSD and, in principle, a fit to the measured C-V curve could be done.
However, for each $U_{B,k}$ three 3d field calculations would need to be performed resulting in 180 field calculations for the whole measured C-V curve. 
Even though each set of three 3d field calculations (with the specified refinement settings)
takes less than a minute on a GPU\footnote{For the calculations presented in this paper the 
following Nvidia GPUs were used: 4 GTX 1080 Ti, 2 Quadro RTX 8000, 6 Tesla V100, 4 RTX 3090 and 2 A100.} in SSD, 
it would not be feasible to set up an optimiser or a Bayesian fit.

Therefore, a deep neural network, $\mathcal{DNN}$, was developed which is able to 
predict the capacitance, $c^{\mathrm{p}}_{12}$, for a set of parameters ($d_{\mathrm{Li}}$, $p_{\zeta_y}$) 
much faster, $\mathcal{O}$(\textmu s).
Here, $U_B$ is added to the input parameters of the model. 
Even though this adds an extra dimension to the model, it simplifies the output as only one capacitance is predicted
instead of a whole C-V curve:
\begin{linenomath}
\begin{align}
    p_{\mathcal{D}} &= (d_{\mathrm{Li}}, p_{\zeta_y}, U_B)\,,\label{eq:DDNparameterDef}\\
    c_{12}^{\mathrm{p}} &= \mathcal{DNN}(p_{\mathcal{D}})\,.\label{eq:DNNcapPrediction}
\end{align}
\end{linenomath}



The impurity model parameters $p_{\mathcal{D}}$ form a 10-dimensional parameter space $\mathcal{P}_{\mathcal{D}}$. The following uniform distributions $\mathcal{U}(a,b)$ were chosen to quasi-randomly draw sets of input parameters for the generation of the training and test capacity datasets:
\begin{linenomath}
\begingroup
\allowdisplaybreaks
\begin{align}
    d_{\mathrm{Li}} &\sim \mathcal{U}(0.2~\textrm{mm}, 5~\textrm{mm})\,,\nonumber\\
    y_1^{\mathrm{bot}}&\sim \mathcal{U}(0.018566145,1)        &&\Leftrightarrow \zeta_{1}^{\mathrm{bot}}\in [\zeta_{l}, \zeta_{\mathrm{u}}]   \,,\nonumber\\
    y_2^{\mathrm{bot}}&\sim \mathcal{U}(0.018566145,1)        &&\Leftrightarrow \zeta_{2}^{\mathrm{bot}}\in [\zeta_{l}, \zeta_{\mathrm{u}}]   \,,\nonumber\\
    y_3^{\mathrm{bot}}&\sim \mathcal{U}(-1,1)                   &&\Leftrightarrow \zeta_{3}^{\mathrm{bot}}\in [-\zeta_{\mathrm{u}}, \zeta_{\mathrm{u}}]  \,,\nonumber\\
    y_4^{\mathrm{bot}}&\sim \mathcal{U}(-1,1)                   &&\Leftrightarrow \zeta_{4}^{\mathrm{bot}}\in [-\zeta_{\mathrm{u}}, \zeta_{\mathrm{u}}]  \,,\nonumber\\
    y_1^{\mathrm{top}}&\sim \mathcal{U}(0.018566145,1)        &&\Leftrightarrow \zeta_{1}^{\mathrm{top}}\in [\zeta_{l}, \zeta_{\mathrm{u}}]   \,,\nonumber\\
    y_2^{\mathrm{top}}&\sim \mathcal{U}(0.018566145,1)        &&\Leftrightarrow \zeta_{2}^{\mathrm{top}}\in [\zeta_{l}, \zeta_{\mathrm{u}}]   \,,\nonumber\\
    y_3^{\mathrm{top}}&\sim \mathcal{U}(-1,1)                   &&\Leftrightarrow \zeta_{3}^{\mathrm{top}}\in [-\zeta_{\mathrm{u}}, \zeta_{\mathrm{u}}]  \,,\nonumber\\
    y_4^{\mathrm{top}}&\sim \mathcal{U}(-1,1)                   &&\Leftrightarrow \zeta_{4}^{\mathrm{top}}\in [-\zeta_{\mathrm{u}}, \zeta_{\mathrm{u}}]  \,,\nonumber\\
    U_B &\sim \mathcal{U}(10~\mathrm{V}, 3000~\mathrm{V}).\nonumber
    \label{eq:DNNparameterSpace}
\end{align}
\endgroup
\end{linenomath}

Note that the distributions of $\zeta_{n}^{\mathrm{bot/top}}$ are not uniform due to the nonlinear transformation defined between $y_{n}^{\mathrm{bot/top}}$ and $\zeta_{n}^{\mathrm{bot/top}}$.

The distribution of $d_{\mathrm{Li}}$ is chosen based on typical lithium layer thickness values.
$U_B$ has to cover all values of $U_{B,k}$.
As Super-Siegfried is an $n$-type germanium detector, the range of the four parameters describing the density
towards the borehole, at the top and bottom, are limited to the $y$ region corresponding to an $n$-type density.
The four parameters describing the density towards the mantle at the top and bottom, however, are allowed to include $p$-type impurity levels.

All individual parameter distributions can easily be transformed to uniform distributions  $\mathcal{U}(0,1)$ on [0, 1]. Therefore, $\mathcal{P}_{\mathcal{D}}$ can be transformed into the 10-dimensional hypercube $[0, 1]^{10}$ with a parameter distribution $\mathcal{U}(0,1)^{10}$.
In order to create a dataset to train the $\mathcal{DNN}$, $N_{\mathrm{s}} = 60000$ samples, were drawn according to
\begin{linenomath}
\begin{equation}
    u_i \sim \mathcal{U}(0,1)^{10} \hspace{1cm}\forall\hspace{0.2cm}i \in \{1, 2, ..., N_{\mathrm{s}}\}\,
\end{equation}
\end{linenomath}
via the quasi-random Golden Sequence~\cite{GoldenSequenceBlockpost:2018,GoldenSequenceJuliaPackage} sampling algorithm.
It generates samples that are very evenly spaced in the unit hypercube.

Machine learning algorithms sometimes fail for parameter spaces with hard edges. 
Therefore, another transformation, $\mathcal{T}_{\mathcal{N}}$, was introduced to transform the $u_i$ into unbound parameter intervals, $[-\infty, \infty]$, such that each of the elements of $u_i^{\mathcal{N}}$ is normally distributed around 0 with a standard deviation of 1:
\begin{linenomath}
\begin{equation}
    u_i^{\mathcal{N}} = \mathcal{T}_{\mathcal{N}}(u_i)\,.
\end{equation}
\end{linenomath}

\begin{figure}[htbp]
    \centering
    \includegraphics[width=\columnwidth]{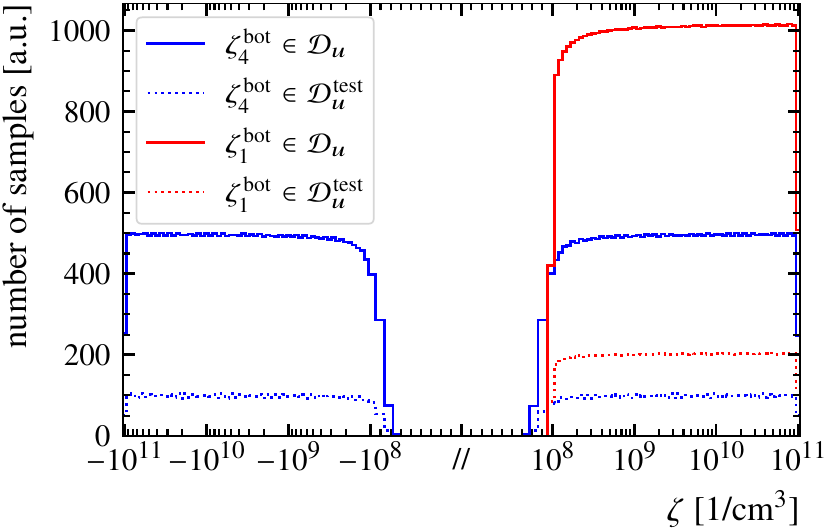}
    \caption{Distributions of $\zeta_{1}^{\mathrm{bot}}$ and $\zeta_{4}^{\mathrm{bot}}$
    from the samples of $\mathcal{D}_u$ and $\mathcal{D}_u^{\mathrm{test}}$.
    }
    \label{fig:histogram_of_generated_samples}
\end{figure}

Each $u_i^{\mathcal{N}}$ still corresponds to a specific $(d_{\mathrm{Li}}, p_{\zeta}, U_B)$.
For all $u_i^{\mathcal{N}}$, the respective $c_{12}^{\mathrm{s}}$ is calculated with SSD to produce a labelled data set,
\begin{linenomath}
\begin{equation}
    \mathcal{D}_u = \left\{(u_i^{\mathcal{N}}| c_{12}^{\mathrm{s},i})\right\} \hspace{0.3cm}\forall\hspace{0.2cm}i \in \{1, 2, ..., N_{\mathrm{s}}\}\,,
\end{equation}
\end{linenomath}
which is divided into a training and a test set with a typical ratio of 80:20,
\begin{linenomath}
\begin{align}
    \mathcal{D}_u^{\mathrm{train}} &= \left\{(u_i^{\mathcal{N}}| c_{12}^{\mathrm{s},i})\right\} \hspace{0.3cm}\forall\hspace{0.2cm}i \in \{1, 2, ..., 0.8\cdot N_{\mathrm{s}}\}\,,\\
    \mathcal{D}_u^{\mathrm{test}}  &= \left\{(u_i^{\mathcal{N}}| c_{12}^{\mathrm{s},i})\right\} \hspace{0.3cm}\forall\hspace{0.2cm}i \in \{0.8\cdot N_{\mathrm{s}}+1, ..., N_{\mathrm{s}}\}\,.
\end{align}
\end{linenomath}
The distribution of generated samples of $\zeta_{1}^{\mathrm{bot}}$ and $\zeta_{4}^{\mathrm{bot}}$ for $\mathcal{D}_u$ and $\mathcal{D}_u^{\mathrm{test}}$
are shown in Fig.~\ref{fig:histogram_of_generated_samples}. 
For both sets, the parameters are properly distributed over the respective parameter intervals.

For machine learning, 
the \emph{Flux.jl}~\cite{Flux:2018,innes:2018} package was used
and the following configuration of $\mathcal{DNN}$ was found to produce good predictions for $c_{12}$:
\begin{itemize}
    \item Number of nodes per layer:\\~\hbox{\hspace{0cm}}[10, 128, 128, 128, 128, 128, 1];
    \item Type of all layers: Dense layer;
    \item Activation function for all but the last layer: GELU;
    \item No activation for the last layer.
\end{itemize}
The function to minimise,
\begin{linenomath}
\begin{equation}
    loss(c_{12}^{\mathrm{p}}, c_{12}^{\mathrm{s}}) = \log(((c_{12}^{\mathrm{p}} - c_{12}^{\mathrm{s}})/\mathrm{pF})^2/h + 1) \cdot h\,,
    \label{eq:DNNLossFunction}
\end{equation}
\end{linenomath}
was chosen as the loss function~\cite{IceCube:2022kff} with $h = 4$.

The training of the $\mathcal{DNN}$ was performed with the samples of $\mathcal{D}_u^{\mathrm{train}}$ 
in 3 subsequent optimisation cycles. The ADAM optimiser algorithm was used with the learning rates $\eta$, number of epochs and batch sizes:
\begin{enumerate}
    \item $\eta = 10^{-2}$, 20 epochs and a batch size of 2048;
    \item $\eta = 10^{-3}$, 80 epochs and a batch size of 1024;
    \item $\eta = 10^{-4}$, 40 epochs and a batch size of 512.
\end{enumerate}

\begin{figure}[htbp]
    \centering
    \includegraphics[width=\columnwidth]{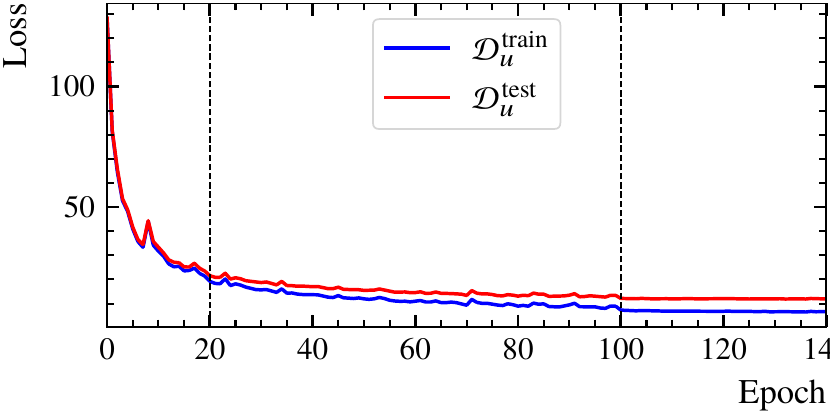}
    \caption{Learning curves of $\mathcal{D}_u^{\mathrm{train}}$ and $\mathcal{D}_u^{\mathrm{test}}$.
     The vertical dashed lines mark the transitions between the three training cycles.}
    \label{fig:DNN_learning_curves}
\end{figure}

The learning curves, i.e. the mean of the loss of all samples of a set after each epoch, of the training and test set
are shown in Fig.~\ref{fig:DNN_learning_curves}.

\begin{figure}[htbp]
    \centering
    \includegraphics[width=\columnwidth]{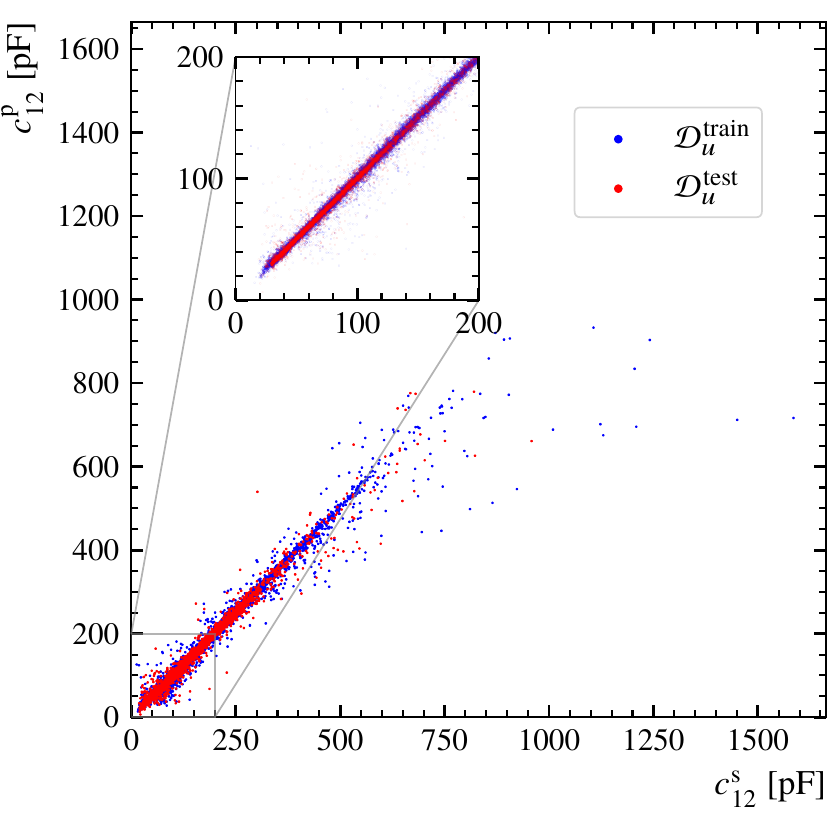}
    \caption{Scatter plot of $c_{12}^{\mathrm{p}}$ over $c_{12}^{\mathrm{s}}$ for
    the samples of $\mathcal{D}_u^{\mathrm{train}}$ and $\mathcal{D}_u^{\mathrm{test}}$ 
    after the training of $\mathcal{DNN}$.}
    \label{fig:DNN_training_result_2d_scatter}
\end{figure}

The predicted capacitances of the trained model over the corresponding "true" value $c_{12}^{\mathrm{s}}$
are shown in Fig.~\ref{fig:DNN_training_result_2d_scatter} as a scatter plot.
A perfect model would produce only points on the diagonal line $c_{12}^{\mathrm{p}} = c_{12}^{\mathrm{s}}$.
Most of the points lie close to that diagonal line and only a few points are further away, so-called outliers.

\begin{figure}[htbp]
    \centering
    \includegraphics[width=\columnwidth]{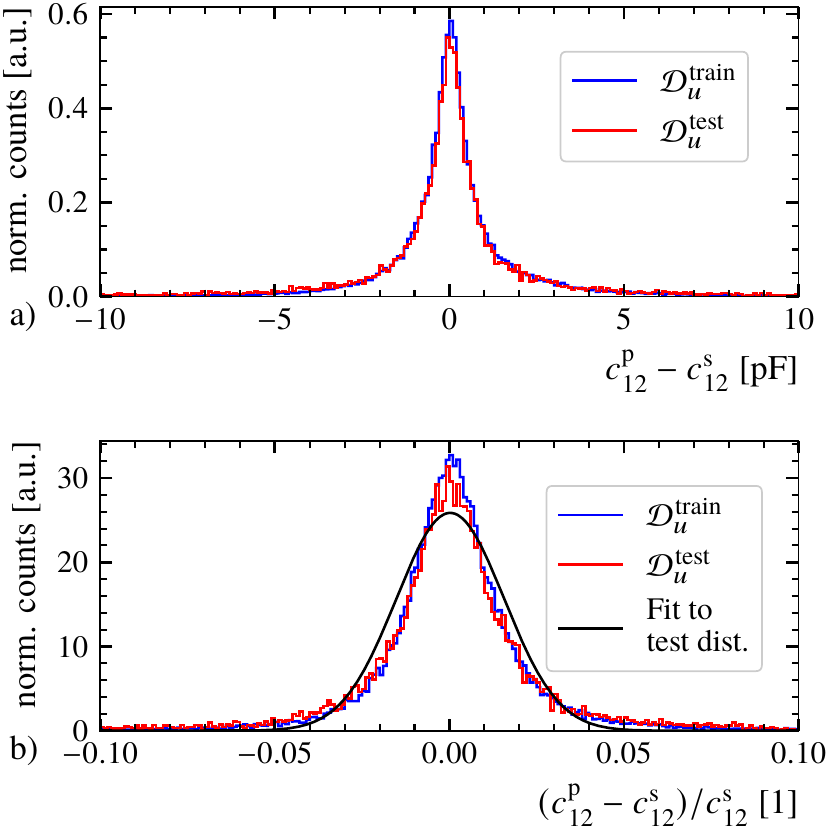}
    \caption{Distribution of a) absolute and b) relative differences between $c_{12}^{\mathrm{p}}$ and $c_{12}^{\mathrm{s}}$ for the samples of $\mathcal{D}_u^{\mathrm{train}}$ and $\mathcal{D}_u^{\mathrm{test}}$.
    A Gaussian approximation of the relative differences of $\mathcal{D}_u^{\mathrm{test}}$ is shown in b).}
    \label{fig:DNN_training_result_1d}
\end{figure}

The distribution of absolute and relative difference between $c_{12}^{\mathrm{p}}$ and $c_{12}^{\mathrm{s}}$
for the training and test set are shown in Fig.~\ref{fig:DNN_training_result_1d}.

A Gaussian approximation is fitted to the distribution of relative differences between $c_{12}^{\mathrm{p}}$ and $c_{12}^{\mathrm{s}}$ for the samples of the test set.
The determined standard deviation of the Gaussian of $1.5\%$ is used to estimate
the uncertainty on the predicted value $c_{12}^{\mathrm{p}}$:
\begin{linenomath}
\begin{equation}
    \sigma_c^{\mathrm{p}}(c_{12}^{\mathrm{p}}) = c_{12}^{\mathrm{p}} \cdot 1.5\%\,.
    \label{eq:CapUncertaintyDNN}
\end{equation}
\end{linenomath}

\section{Bayesian fits of the impurity density}
\label{sec:BayesianFits}

A predicted C-V curve $c_{12}^{\mathrm{p},k}$ can now be produced for a given set of $(d_{\mathrm{Li}}, p_{\zeta})$ by evaluating 
the trained network $\mathcal{DNN}$ at $(d_{\mathrm{Li}}, p_{\zeta}, U_{B,k})$ for all $k$.

For each predicted $c_{12}^{\mathrm{p}}$, an uncertainty is estimated by Gaussian error propagation
of the three different sources of uncertainty: 
\begin{linenomath}
\begin{equation}
    \sigma_c(c_{12}^{\mathrm{p}})^2 = 
        \sigma_c^{\mathrm{m}}(c_{12}^{\mathrm{p}})^2
      + \sigma_c^{\mathrm{s}}(c_{12}^{\mathrm{p}})^2
      + \sigma_c^{\mathrm{p}}(c_{12}^{\mathrm{p}})^2\,,
    \label{eq:ModelUncertainty}
\end{equation}
\end{linenomath}
where $\sigma_c^{\mathrm{s}}$ is the uncertainty due to SSD, see Eq.\eqref{eq:CapUncertaintySSD}, and
$\sigma_c^{\mathrm{p}}$ is the uncertainty due to $\mathcal{DNN}$, see Eq.\eqref{eq:CapUncertaintyDNN}.
The uncertainty due to the measurements, $\sigma_c^{\mathrm{m}}$, is motivated
by the conservatively estimated uncertainty on $c_{12}^{\mathrm{m}}$, 
see Sec.~\ref{sec:MeasuredCVcurve}:
\begin{linenomath}
\begin{equation}
    \sigma_c^{\mathrm{m}}(c_{12}^{\mathrm{p}}) = c_{12}^{\mathrm{p}} \cdot 3\%\,.
    \label{eq:CapUncertaintyData}
\end{equation}
\end{linenomath}
The likelihood $\mathcal{L}$ of a predicted C-V curve $c_{12}^{\mathrm{p},k}$
can then be defined as
\begin{linenomath}
\begin{equation}
    \mathcal{L}(c_{12}^{\mathrm{p},k}) = \prod_{k = 1}^{60}\mathcal{N}\left(c_{12}^{\mathrm{p},k}, 
                                              \sigma_c(c_{12}^{\mathrm{p},k})\right)(c_{12}^{\mathrm{m},k})\,,
    \label{eq:Likelihood}
\end{equation}
\end{linenomath}
which is the product over all $k$ of normal distributions with mean value $c_{12}^{\mathrm{p},k}$
and standard deviation $\sigma_c(c_{12}^{\mathrm{p},k})$ evaluated at $c_{12}^{\mathrm{m},k}$.

The software package \emph{BAT.jl}~\cite{Schulz:2021BAT} was used to perform Bayesian fits for two different cases 
of $\zeta_{\mathrm{RZ}}$: One without any radial dependence of the impurity density, $\mathcal{B}_{\mathrm{Z}}$, and
one with a radial dependence through all 8 parameters of $p_{\zeta}$, $\mathcal{B}_{\mathrm{RZ}}$.

\subsection{Bayesian fit of the impurity density without radial dependence}
\label{sec:BayesianFitNoRadialModulation}

For $\mathcal{B}_{\mathrm{Z}}$, the model for the fit only has 3 parameters:
\begin{linenomath}
\begin{equation}
    p_{\mathcal{B}_{\mathrm{Z}}} = (d_{\mathrm{Li}}, \zeta^{\mathrm{bot}}, \zeta^{\mathrm{top}})\,,
\end{equation}
\end{linenomath}
where the two parameters for the impurity density level are used 
at all $r_{b,i}$ at the top and bottom in the $\zeta_{\mathrm{RZ}}$ model, respectively. 
A flat prior was chosen for $d_{\mathrm{Li}}$. For 
$\zeta^{\mathrm{bot}}$ and $\zeta^{\mathrm{top}}$, 
the prior distributions were chosen as broad truncated\footnote{
All truncated distributions presented in this paper are cut off at the endpoints
of the interval of the respective parameter space.} normal distributions, $\mathcal{N}_t$, 
in $y$ around the respective $y$ for
the impurity levels provided by the manufacturer $\zeta_{\mathrm{M}}^{\mathrm{bot}}$ and $\zeta_{\mathrm{M}}^{\mathrm{top}}$.
The broadness was chosen to express the large uncertainty on the values provided by the 
manufacturer and such that the entire parameter space was tested.
The prior distributions for all three parameters are listed in Tab.~\ref{tab:Prior_B_Z}.

\begin{table}[htbp]
    \centering
    \caption{Prior distribution for all parameters from $\mathcal{B}_{\mathrm{Z}}$.
    See main text for reasoning.}
    \begin{normalsize}
    \begin{tabular}{c|l}
         & Prior distribution\\
        \hline
        $d_{\mathrm{Li}}$       & $d_{\mathrm{Li}} \hspace{0.09cm} \sim \mathcal{U}(0.2~\textrm{mm}, 5~\textrm{mm})$ \\
        $\zeta^{\mathrm{bot}}$  & $y^{\mathrm{bot}} \sim \mathcal{N}_t(0.7062704, 0.4)$\\
        $\zeta^{\mathrm{top}}$ & $y^{\mathrm{top}} \sim \mathcal{N}_t(0.55060154, 0.4)$\\
    \end{tabular}
    \end{normalsize}
    \label{tab:Prior_B_Z}
\end{table}

\begin{figure}[htbp]
    \centering
    \includegraphics[width=\columnwidth]{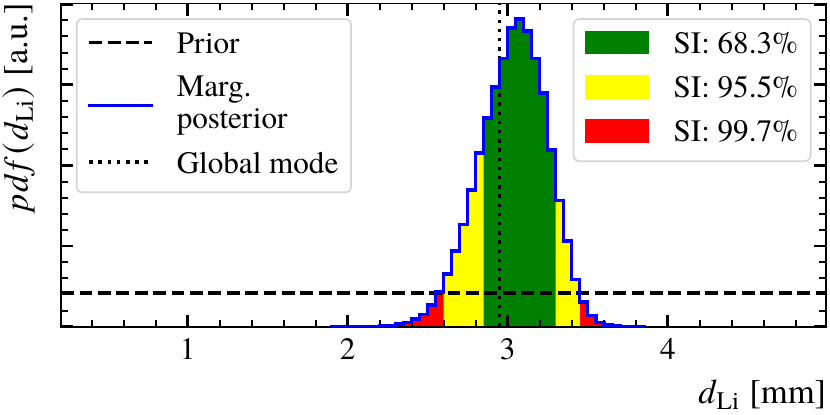}
    \caption{Prior distribution and marginalised posterior distribution of $d_{\mathrm{Li}}$ 
        from $\mathcal{B}_{\mathrm{Z}}$. 
        The smallest intervals (SI) containing certain amounts of probability are shown as shaded areas.}
    \label{fig:bat_fit_prior_and_mPosterior_linear_LiThickness}
\end{figure}

The marginalised posterior distribution and the prior distribution of $d_{\mathrm{Li}}$ is shown in
Fig.~\ref{fig:bat_fit_prior_and_mPosterior_linear_LiThickness}.
The fit indicates a thickness of about 3\,mm, which is quite thick.
However, this parameter of the model is mainly sensitive to the end of the C-V curve
and probably heavily impacted by possible imperfections of the implementation of the geometry 
of the detector.
In addition, the detector is old and some growth of $d_{\mathrm{Li}}$ is expected.

\begin{figure}[htbp]
    \centering
    \includegraphics[width=\columnwidth]{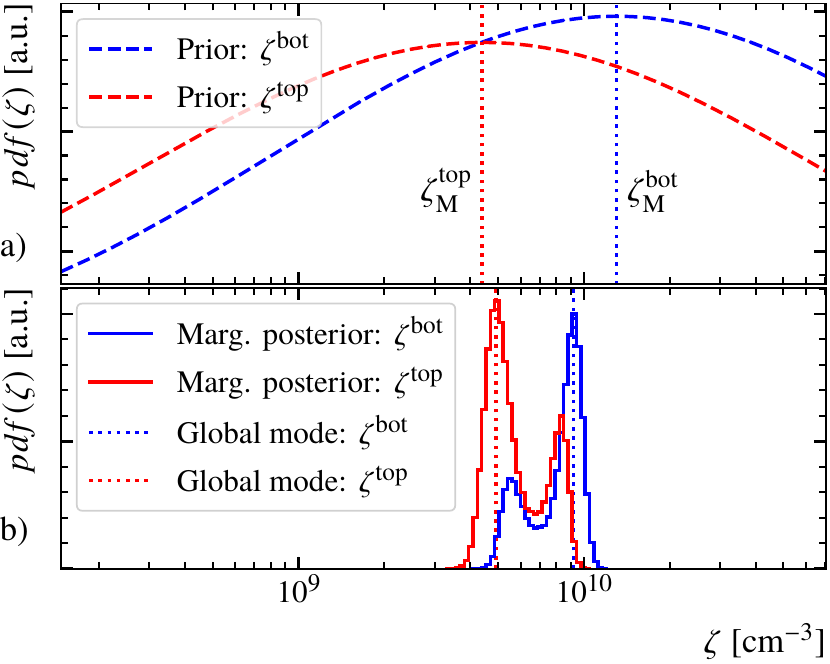}
    \caption{a) Prior distributions and b) marginalised posterior distributions 
    for $\zeta^{\mathrm{bot}}$ and $\zeta^{\mathrm{top}}$ from $\mathcal{B}_{\mathrm{Z}}$.}
    \label{fig:bat_fit_prior_and_mPosterior_linear_ImpLevels}
\end{figure}

The prior distributions and marginalised posterior distributions 
of $\zeta^{\mathrm{bot}}$ and $\zeta^{\mathrm{top}}$
are shown in Fig.~\ref{fig:bat_fit_prior_and_mPosterior_linear_ImpLevels}.

\begin{figure}[htbp]
    \centering
    \includegraphics[width=\columnwidth]{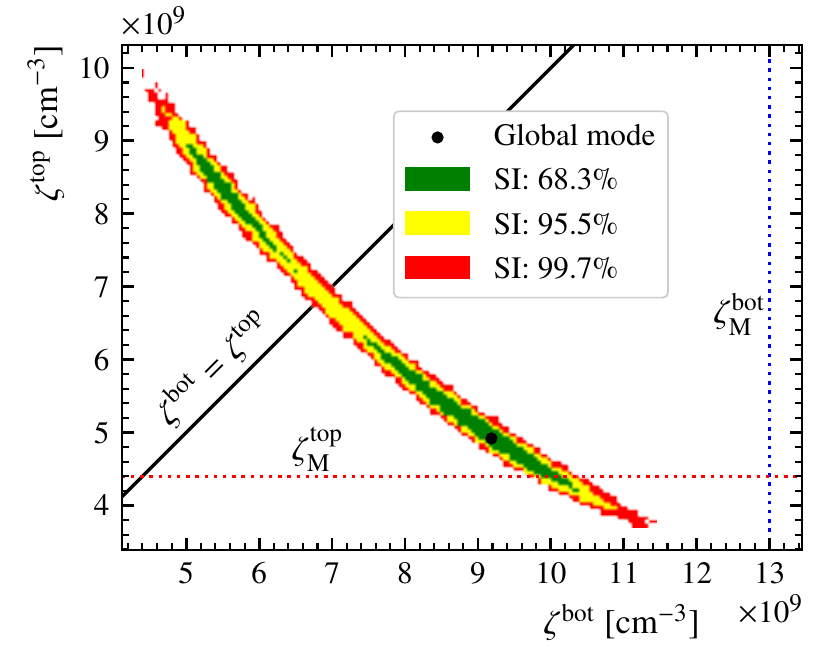}
    \caption{2d-marginalised posterior distribution of $\zeta^{\mathrm{top}}$ and $\zeta^{\mathrm{bot}}$ 
        of $\mathcal{B}_{\mathrm{Z}}$. 
        The smallest intervals (SI) containing certain amounts of probability are shown as shaded areas.}
    \label{fig:bat_fit_2D_Posterior_linear_ImpLevels}
\end{figure}

The posterior distributions peaks are not too far away from $\zeta_{\mathrm{M}}^{\mathrm{top}}$ and $\zeta_{\mathrm{M}}^{\mathrm{bot}}$. 
However, there are two modes in both posterior distributions.
This is also visible in the 2d-marginalised posterior distribution of $\zeta^{\mathrm{bot}}$ and $\zeta^{\mathrm{top}}$ in Fig.~\ref{fig:bat_fit_2D_Posterior_linear_ImpLevels}.
The second mode can be explained by the symmetry of the setup. The detector could basically be 
physically inverted such that the top and bottom would be switched.

\begin{figure}[htbp]
    \centering
    \includegraphics[width=\columnwidth]{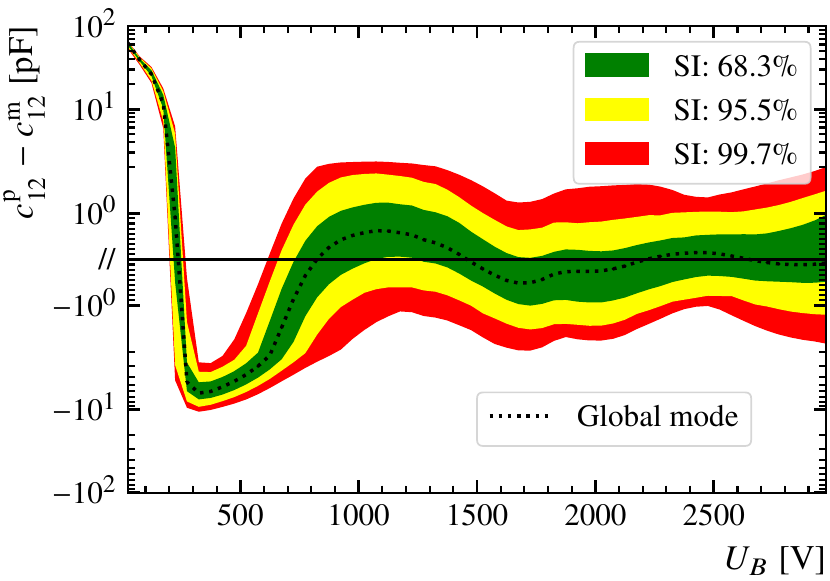}
    \caption{Posterior predictive of the difference between the predicted, $c_{12}^{\mathrm{p}}$, 
        and measured, $c_{12}^{\mathrm{m}}$, capacitances from $\mathcal{B}_{\mathrm{Z}}$.
        The shaded bands mark areas of how probable
        a value of $c_{12}^{\mathrm{p}}-c_{12}^{\mathrm{m}}$ 
        is based on the posterior of~$\mathcal{B}_{\mathrm{Z}}$.}
    \label{fig:bat_fit_posterior_predictive_linear}
\end{figure}

\begin{figure}[htbp]
    \centering
    \includegraphics[width=\columnwidth]{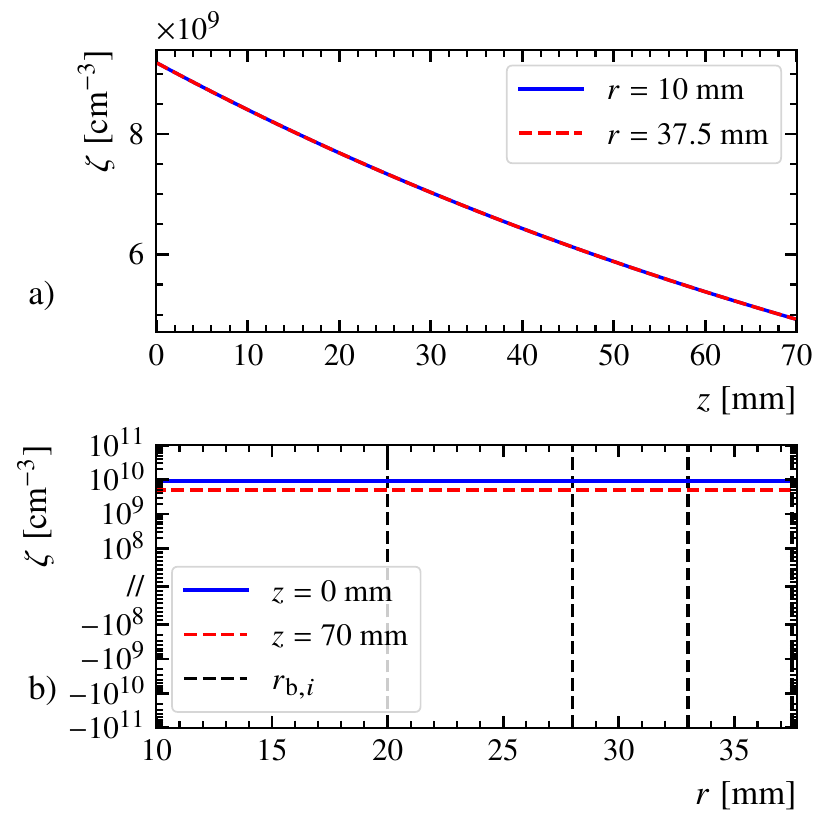}
    \caption{$\zeta_{\mathrm{RZ}}$ for the global mode from $\mathcal{B}_{\mathcal{Z}}$ as a function of a) $z$ and b)~$r$.}
    \label{fig:fitted_impurity_density_linear}
\end{figure}

The posterior predictive of the difference between the predicted and measured capacitances, 
$c_{12}^{\mathrm{p}} - c_{12}^{\mathrm{m}}$, is shown in Fig.~\ref{fig:bat_fit_posterior_predictive_linear}.
At voltages above 1000\,V the posterior predictive is centred around 0. However, below 1000\,V, it becomes very clear that the model with no radial dependence is not able to 
describe the measured C-V curve as the predictions do not describe the measurements.
Lower bias voltages correspond to larger radii as the depleted region 
grows from the mantle towards the borehole, see Fig.~\ref{fig:ssd_impM_point_types}.
Thus, a radial dependence of the impurities towards larger radii is again suggested.
The impurity density $\zeta_{\mathrm{RZ}}$ for the global mode of the fit $\mathcal{B}_{\mathrm{Z}}$
is shown in Fig.~\ref{fig:fitted_impurity_density_linear}.

\subsection{Bayesian fit of the impurity density with radial dependence}
\label{sec:BayesianFitWithRadialModulation}

For $\mathcal{B}_{\mathrm{RZ}}$, the model for the fit has 9 parameters:
\begin{linenomath}
\begin{equation}
    p_{\mathcal{B}_{\mathrm{RZ}}} = (d_{\mathrm{Li}}, p_{\zeta})\,.
\end{equation}
\end{linenomath}

The prior distributions for all parameters are shown in Tab.~\ref{tab:Prior_B_RZ}.
For $d_{Li}$, a flat prior was chosen.
For the 6 parameters $\zeta_{1-3}^{\mathrm{top/bot}}$,
truncated normal distributions based on the marginalised posterior distributions 
of $\zeta^{\mathrm{bot}}$ and $\zeta^{\mathrm{top}}$ from $\mathcal{B}_{\mathrm{Z}}$ were chosen.
For the two parameters $\zeta_{4}^{\mathrm{top/bot}}$, broad truncated normal distributions centred around 0
were chosen based on the conclusions drawn from $\mathcal{B}_{\mathrm{Z}}$,
that assumed impurities in $\mathcal{B}_{\mathrm{Z}}$ were too high at larger radii.

\begin{table}[htbp]
    \centering
    \caption{Prior distribution for all parameters from $\mathcal{B}_{\mathrm{RZ}}$.
    See main text for reasoning.}
    \begin{normalsize}
    \begin{tabular}{c|l}
         & Prior distribution\\
        \hline
        $d_{\mathrm{Li}}$         & $d_{\mathrm{Li}}\hspace{0.09cm} \sim \mathcal{U}(0.2~\textrm{mm}, 5~\textrm{mm})$ \\
        $\zeta_1^{\mathrm{bot}}$  & $y_1^{\mathrm{bot}} \sim \mathcal{N}_t(0.6563343, 0.1)$\\
        $\zeta_2^{\mathrm{bot}}$  & $y_2^{\mathrm{bot}} \sim \mathcal{N}_t(0.6563343, 0.1)$\\
        $\zeta_3^{\mathrm{bot}}$  & $y_3^{\mathrm{bot}} \sim \mathcal{N}_t(0.6563343, 0.1)$\\
        $\zeta_4^{\mathrm{bot}}$  & $y_4^{\mathrm{bot}} \sim \mathcal{N}_t(0, 1.0)$\\    
        $\zeta_1^{\mathrm{top}}$  & $y_1^{\mathrm{top}} \sim \mathcal{N}_t(0.56663054, 0.1)$\\
        $\zeta_2^{\mathrm{top}}$  & $y_2^{\mathrm{top}} \sim \mathcal{N}_t(0.56663054, 0.1)$\\
        $\zeta_3^{\mathrm{top}}$  & $y_3^{\mathrm{top}} \sim \mathcal{N}_t(0.56663054, 0.1)$\\
        $\zeta_4^{\mathrm{top}}$  & $y_4^{\mathrm{top}} \sim \mathcal{N}_t(0, 1.0)$\\
        
    \end{tabular}
    \end{normalsize}
    \label{tab:Prior_B_RZ}
\end{table}

\begin{figure}[htbp]
    \centering
    \includegraphics[width=\columnwidth]{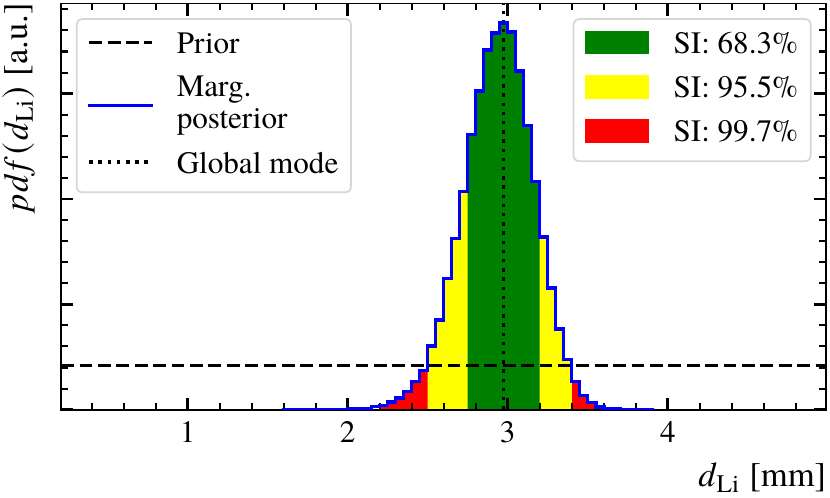}
    \caption{Prior distribution and marginalised posterior distribution of $d_{\mathrm{Li}}$ 
        from $\mathcal{B}_{\mathrm{RZ}}$.
        The smallest intervals (SI) containing certain amounts of probability are shown as shaded areas.}
    \label{fig:bat_fit_prior_and_mPosterior_linear_and_radial_LiThickness}
\end{figure}

The prior distribution and marginalised posterior distribution of $d_{\mathrm{Li}}$ are shown in 
Fig.~\ref{fig:bat_fit_prior_and_mPosterior_linear_and_radial_LiThickness}. 
The posterior distribution is very similar to the case without radial dependence.

\begin{figure*}[htbp]
    \centering
    \includegraphics[width=\textwidth]{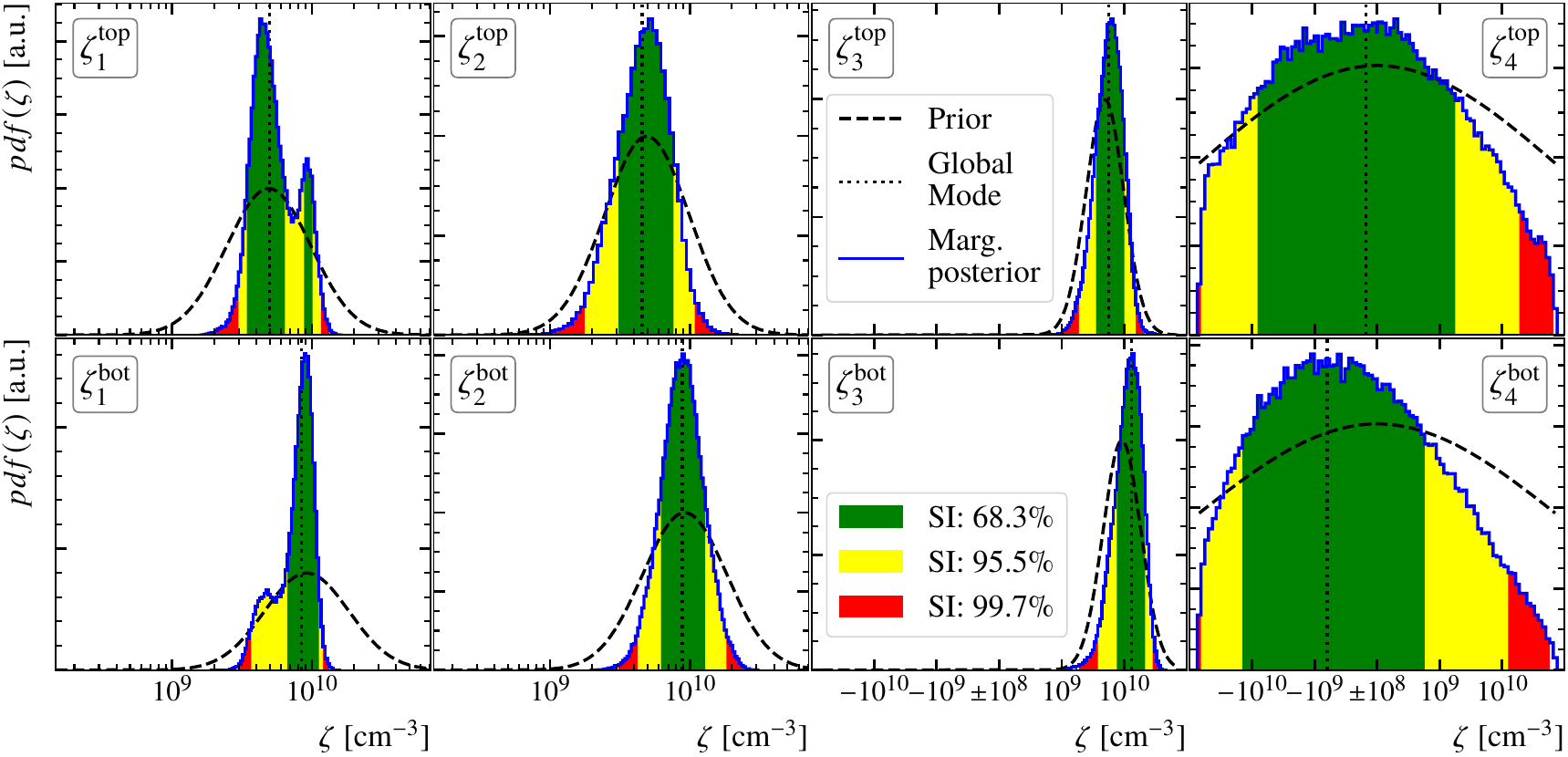}
    \caption{Prior distributions and marginalised posterior densities of the 8 parameters describing $\zeta_{\mathrm{RZ}}$ of $\mathcal{B}_{\mathrm{RZ}}$. 
    The $\zeta$-axes cover the total parameter space for all parameters.
    The smallest intervals (SI) containing certain amounts of probability are shown as shaded areas.}
    \label{fig:bat_fit_marg_posterior_imp_levels_linear_and_radial}
\end{figure*}

The prior distributions and marginalised posterior distributions for the 8 parameters of $\zeta_{1-4}^{\mathrm{top/bot}}$
are shown in Fig.~\ref{fig:bat_fit_marg_posterior_imp_levels_linear_and_radial}.

The two posterior distributions at $r_{b,1}$ are very similar to the posterior distributions
of the case without radial dependence. That was expected since the posterior predictive of
the first case indicated that no modulation is required at smaller radii
but is required at larger radii.
This is also indicated by the posterior distribution of the two parameters $\zeta_{4}^{\mathrm{top/bot}}$
which describe the impurity density at the mantle of the detector.
The two posterior distributions are very broad, but favour a very low impurity density and even a 
possible $p$-type volume close to the mantle. 
The broadness comes from the increasing estimated relative uncertainty on the predicted capacitances towards 
lower bias voltages because the capacitance becomes in general larger towards lower bias voltages.
Another reason for the broadness is that the two parameters are only sensitive 
to the first few data points of the measured C-V curve.

\begin{figure}[htbp]
    \centering
    \includegraphics[width=\columnwidth]{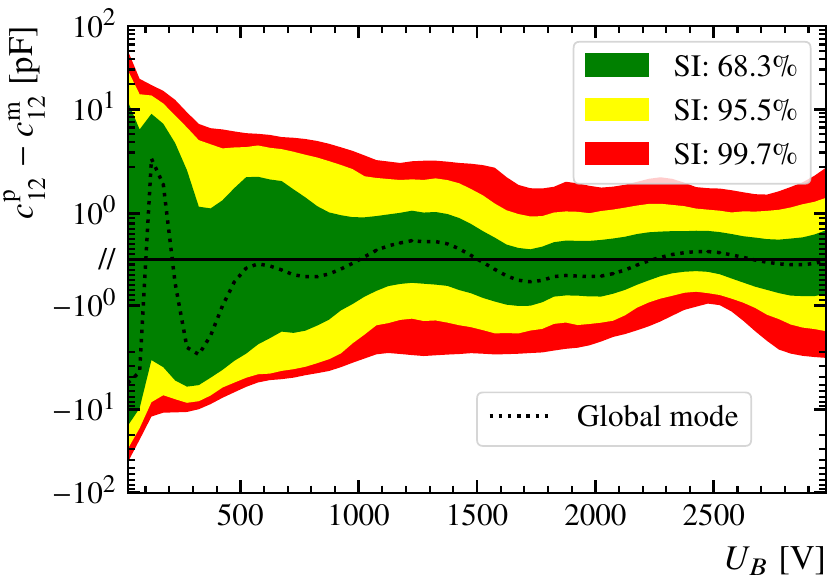}
    \caption{Posterior predictive of the difference between the predicted and measured capacitances 
        of $\mathcal{B}_{\mathrm{RZ}}$.
        The shaded bands mark areas of how probable
        a value of $c_{12}^{\mathrm{p}}-c_{12}^{\mathrm{m}}$ 
        is based on the posterior of~$\mathcal{B}_{\mathrm{RZ}}$.}
    \label{fig:bat_fit_posterior_predictive_linear_and_radial}
\end{figure}

The posterior predictive of the difference between the predicted and measured capacitances, 
$c_{12}^{\mathrm{p}} - c_{12}^{\mathrm{m}}$, is shown in Fig.~\ref{fig:bat_fit_posterior_predictive_linear_and_radial}.

In contrast to the $\mathcal{B}_{\mathrm{Z}}$ model, the bands are centred around zero for all bias voltages.
The predictions become less precise towards lower bias voltages due to the previously explained reasons for the broadness of 
the marginalised posterior distributions of $\zeta_{4}^{\mathrm{top}}$ and $\zeta_{4}^{\mathrm{bot}}$.

\begin{figure}[htbp]
    \centering
    \includegraphics[width=\columnwidth]{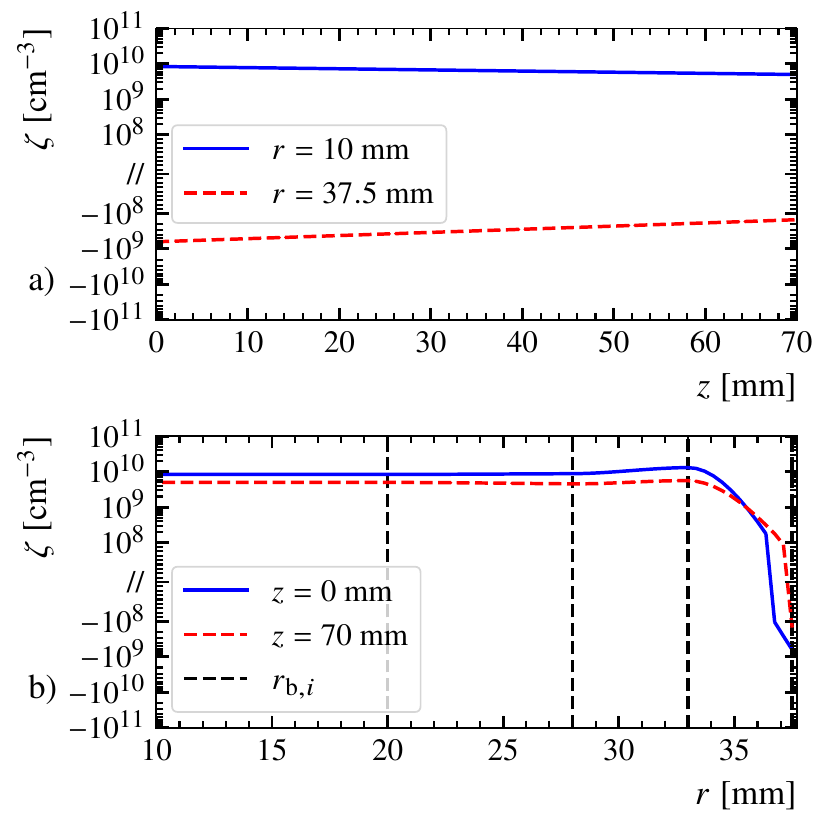}
    \caption{Global mode of the fitted impurity density of $\mathcal{B}_{\mathrm{RZ}}$.}
    \label{fig:fitted_impurity_density_linear_and_radial}
\end{figure}

The impurity density $\zeta_{\mathrm{RZ}}$ for the global mode of the fit is shown in Fig.~\ref{fig:fitted_impurity_density_linear_and_radial}.

\section{Impact of impurity densities on pulse formation}
\label{sec:Discussion}

The electric field, $\mathcal{\vec{E}}$, and the pulses of one event spawned 
at (r = 37\,mm, $\varphi$ = 30$^{\circ}$, z = 35\,mm) were simulated with SSD for three different 
$\zeta$ at the operation voltage of 3000\,V:
\begin{itemize}
    \item $\zeta_{\mathrm{M}}$ with $d_{\mathrm{Li}} = 3$\,mm: $\zeta_{\mathrm{M}}^*$\footnote{This was done in order to have approximately the same $d_{\mathrm{Li}}$ for all models.}\,,
    \item $\zeta_{\mathrm{RZ}}$ for the global mode of $\mathcal{B}_{\mathrm{Z}}$:
        $\zeta_{\mathcal{B}_{\mathrm{Z}}}$\,,
    \item $\zeta_{\mathrm{RZ}}$ for the global mode of $\mathcal{B}_{\mathrm{RZ}}$:
        $\zeta_{\mathcal{B}_{\mathrm{RZ}}}$\,.
\end{itemize}

\begin{figure}[htbp]
    \centering
    \includegraphics[width=\columnwidth]{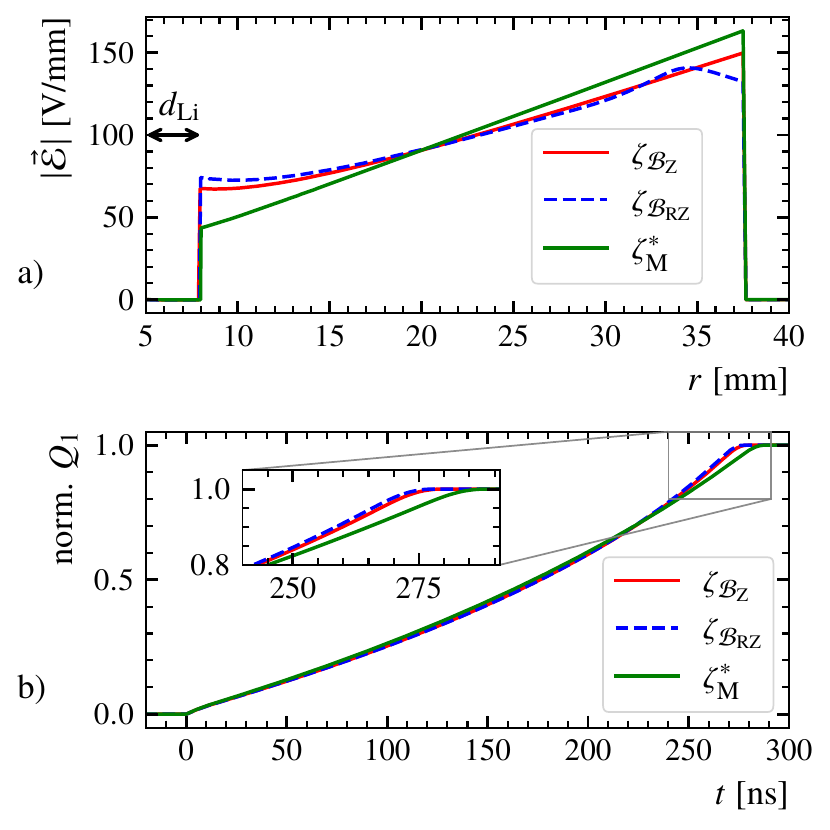}
    \caption{a) The electric field strength at ($\varphi$ = 30$^{\circ}$, z = 35\,mm) over $r$ and
    b) normalised pulses of the $n^+$ contact of an event spawned 
    at (r = 37\,mm, $\varphi$ = 30$^{\circ}$, z = 35\,mm) 
    as simulated with SSD for the three different impurity density distributions 
    $\zeta_{\mathcal{B}_{\mathrm{Z}}}$, $\zeta_{\mathcal{B}_{\mathrm{RZ}}}$ and $\zeta_{\mathrm{M}}^*$.
    }
    \label{fig:field_and_pulse_comparisons}
\end{figure}

The electric field strength at ($\varphi$ = 30$^{\circ}$, z = 35\,mm) over $r$ 
and the normalised pulses from the $n^+$ contact of the simulated event 
are shown in Fig.~\ref{fig:field_and_pulse_comparisons} for the three different $\zeta$ models.

The electric field strength close to the contacts 
differs significantly for the three cases.
The field strength from $\zeta_{\mathcal{B_{\mathrm{Z}}}}$ 
is significantly less radius dependent than that
from $\zeta_{\mathrm{M}}^*$. 
This causes the pulse to become faster at the end.
The radial decrease of impurities in $\zeta_{\mathcal{B_{\mathrm{RZ}}}}$ 
further reduces $|\mathcal{\vec{E}}|$ close to $p^+$ contact.
The field strength close to the $n^+$ is further increased.
Nevertheless, the pulses for $\zeta_{\mathcal{B}_{\mathrm{Z}}}$ and
$\zeta_{\mathcal{B}_{\mathrm{RZ}}}$ are very similar.

However, the effect of $\zeta$ on the simulated pulses also depends on the 
charge drift model describing the mobility tensor 
and its dependence on the electric field.
For the simulated pulses shown in Fig.~\ref{fig:field_and_pulse_comparisons}b, 
the charge drift model from the AGATA Detector Library~\cite{Bruyneel:2006764, Bruyneel:2016zih} as implemented in SSD~\cite{Abt:2021mzq}
was used with its default parameters.
The differences in the pulses for the three $\zeta$ models show 
how important it is to use the correct impurity density distribution 
when using pulse shapes from measurements to tune the parameters
of the drift model.

\section{Summary and Outlook}
\label{sec:Summary}

The capacitance matrix of a germanium detector was explained in detail and
it was shown how the capacitances depend on the depletion of the detector and, thus,
on the impurity density distribution of the crystal. 
The setup K1 and the true-coaxial $n$-type germanium detector Super-Siegfried
were introduced and it was explained how to measure one of the elements 
of the capacitance matrix of the detector for different bias voltages.
The measured C-V curve was compared to a C-V curve simulated 
for the impurity density distribution as provided by manufacturer.
The comparison suggested a radial dependence of impurity densities.
This was confirmed by a Bayesian fit which optimised the impurity 
density model with only a dependence on the $z$-axis of the detector.
A model including a radial dependence of the impurity density was introduced.
The Bayesian fit of this model to the measured C-V curve
provided a good description of the data.
This indicates that the crystal under study really has an $r$ dependent 
impurity density distribution with a very low level of 
electrically active impurities close to the detector edge.

A novel method was introduced that uses a deep neural network, 
trained on GPU-accelerated capacitance calculations, 
to enable full Bayesian parameter inference on complex impurity density models.

The possibility to determine impurity density distributions from 
capacitance measurements opens a road to study mobility tensors and drift models
by comparing measured and simulated pulses without the uncertainties otherwise
introduced by the lack of knowledge on these impurity densities.
The knowledge of the impurity densities is also important for pulse-shape
analysis used in rare-event searches where the exact understanding of 
the pulse formation is critical to discriminate between signal and background events.

It should be noted that the method presented here can also be used to optimise 
general detector properties during the detector design phase. In addition, 
the method has the potential to determine impurity distributions based on impurity-sensitive 
detector properties other than capacitance. Inferring impurity from voltage-dependent properties 
like the shape of the depletion volume, determined by Compton scanning, 
or the total active volume will be the subject of future work.

\bibliographystyle{spphys} 

\bibliography{main.bib}

\end{document}